\documentclass[pdflatex,sn-mathphys-num]{sn-jnl}% Math 

\usepackage{graphicx}%
\usepackage{multirow}%
\usepackage{amsmath,amssymb,amsfonts}%
\usepackage{amsthm}%
\usepackage{mathrsfs}%
\usepackage[title]{appendix}%
\usepackage{xcolor}%
\usepackage{textcomp}%
\usepackage{manyfoot}%
\usepackage{booktabs}%
\usepackage{algorithm}%
\usepackage{algorithmicx}%
\usepackage{algpseudocode}%
\usepackage{listings}%
\usepackage{soul}
\usepackage{verbatim}
\usepackage{amssymb}
\usepackage{chngcntr}

\usepackage[font=small]{caption}  
\theoremstyle{thmstyleone}

\theoremstyle{thmstyletwo}%
\theoremstyle{thmstylethree}%
\begin{document}
\title[Article Title]{All-in-One Analog AI Hardware: On-Chip Training and Inference with Conductive-Metal-Oxide/HfO\textsubscript{\textnormal{x}} ReRAM Devices}

\author*[1]{\fnm{Donato Francesco} \sur{Falcone}}\email{dof@zurich.ibm.com}
\author[1]{\fnm{Victoria} \sur{Clerico}}
\author[1]{\fnm{Wooseok} \sur{Choi}}
\author[1]{\fnm{Tommaso} \sur{Stecconi}}
\author[1]{\fnm{Folkert} \sur{Horst}}
\author[1]{\fnm{Laura} \sur{Bégon-Lours}}
\author[1]{\fnm{Matteo} \sur{Galetta}}
\author[1]{\fnm{Antonio} \sur{La Porta}}
\author[2,3]{\fnm{Nikhil} \sur{Garg}}
\author[2,3]{\fnm{Fabien} \sur{Alibart}}
\author[1]{\fnm{Bert Jan} \sur{Offrein}}
\author[1]{\fnm{Valeria} \sur{Bragaglia}}

\affil[1]{\orgname{IBM Research - Europe}, \orgaddress{\city{Rüschlikon}, \postcode{8803}, \state{Zürich}, \country{Switzerland}}}

\affil[2]{\orgdiv{Institut Interdisciplinaire d’Innovation Technologique (3IT)}, \orgname{Université de Sherbrooke}, \orgaddress{\city{Sherbrooke}, \postcode{QC J1K 0A5}, \state{Quebec}, \country{Canada}}}

\affil[3]{\orgdiv{Institute of Electronics, Microelectronics and Nanotechnology (IEMN)}, \orgname{Université de Lille}, \orgaddress{\city{Villeneuve d’Ascq}, \postcode{59650}, \country{France}}}

% ALREADY DONE compared to submission in AFM
%TO DO: CAMBIARE 512X512 IN 512X512
%TO DO: MODIFICARE LA FORMULA CHE PUO ESSERE MISLEADING IN FIG. 3E, vedi la TESI
% CHANGED Y-AXIS in Fig4a to probability density

% BETTER TO DO AFTER REVISION/ Be careful
% Be aware --> In Fig. 4e, normalizzi ogni guassiana, ecco perche hai sempre picco 1 e area maggiore per visualizzazione. Quindi la normalizzazione e' fatta ad ogni time. Potresti scriverlo tipo for graphical ..each guassian is normalized.
\abstract{Analog in-memory computing is an emerging paradigm designed to efficiently accelerate deep neural network workloads. Recent advancements have focused on either inference or training acceleration. However, a unified analog in-memory technology platform—capable of on-chip training, weight retention, and long-term inference acceleration—has yet to be reported. This work presents an all-in-one analog AI accelerator, combining these capabilities to enable energy-efficient, continuously adaptable AI systems. The platform leverages an array of analog filamentary conductive-metal-oxide (CMO)/HfO\textsubscript{\textnormal{x}} resistive switching memory cells (ReRAM) integrated into the back-end-of-line (BEOL). The array demonstrates reliable resistive switching with voltage amplitudes below 1.5~V, compatible with advanced technology nodes. The array’s multi-bit capability (over 32 stable states) and low programming noise (down to 10~nS) enable a nearly ideal weight transfer process, more than an order of magnitude better than other memristive technologies. Inference performance is validated through matrix-vector multiplication simulations on a 64×64 array, achieving a root-mean-square error improvement by a factor of 20 at 1 second and 3 at 10 years after programming, compared to state-of-the-art. Training accuracy closely matching the software equivalent is achieved across different datasets. The CMO/HfO\textsubscript{\textnormal{x}} ReRAM technology lays the foundation for efficient analog systems accelerating both inference and training in deep neural networks.}

\keywords{In-memory computing, Analog ReRAM, Deep Neural Networks, Training, Inference}

\maketitle
\section{Introduction}\label{sec1}
\begin{figure}[H]
\centering
\includegraphics[width=1.0\textwidth]{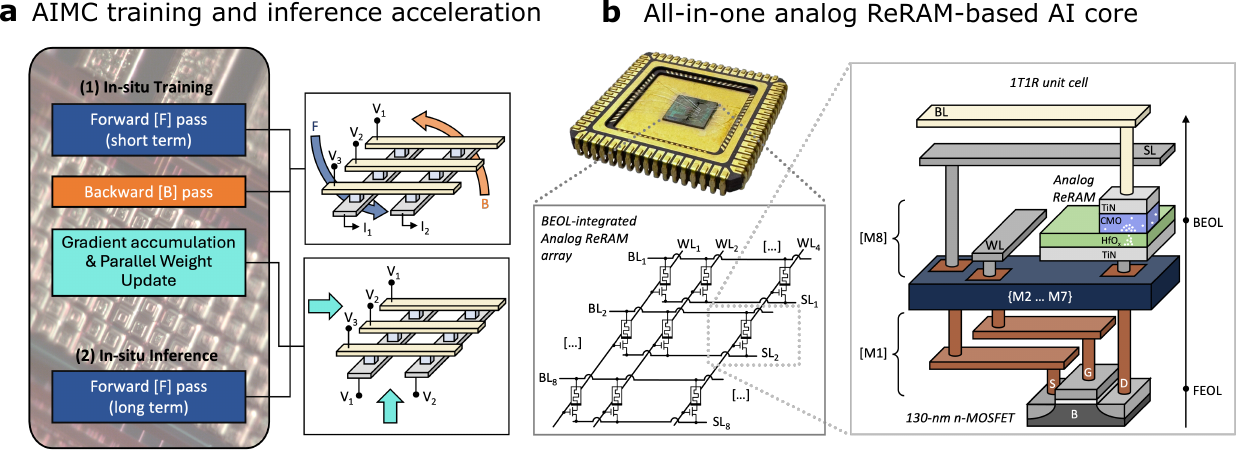}
\caption{\textbf{All-in-one AIMC challenge.} \textbf{a} Schematic representation of the key steps required to perform on-chip training and inference with analog acceleration. Each step is executed using a crossbar array of resistive devices. \textbf{b} CMO/HfO\textsubscript{\textnormal{x}} ReRAM AI core used in this work, consisting of an 8×4 array of 1T1R unit cells. From a fabrication perspective, each ReRAM cell is integrated into the BEOL of a \(\mathrm{130 \, nm}\) NMOS transistor with copper interconnects.}
\label{fig1}
\end{figure}

Modern computing systems rely on von Neumann architectures, where instructions and data must be transferred between memory and the processing unit to perform computational tasks. This data transfer, particularly recurrent and massive in prominent artificial intelligence (AI)-related workloads, results in significant latency and energy overhead \cite{epoch2024datamovementbottlenecksscalingpast1e28flop}. Digital AI accelerators address this challenge through computational parallelism, bringing memory closer to the processing units, and exploiting application-specific processors \cite{Jouppi2017,Sze2017}. This approach has demonstrated to bring significant improvements in throughput and efficiency for running deep neural networks (DNNs) \cite{Haensch2019}, but the physical separation between memory and compute units persists. Analog in-memory computing (AIMC) \cite{Sebastian2020} is a promising approach to eliminate this separation and so achieve further power and efficiency improvements in deep-learning workloads \cite{Mutlu2019}, by enabling some arithmetic and logic operations to be performed directly at the location where the data is stored. By mapping the weights of DNNs onto crossbar arrays of resistive devices and by leveraging Ohm's and Kirchhoff’s physical laws, matrix-vector multiplications (MVMs)—the most recurrent operation in AI-workloads \cite{Tsai2023}—are performed in memory with $O(1)$ time complexity \cite{Sebastian2020,Burr2017,Haensch2019}. Recent demonstrations of the AIMC paradigm have primarily focused on accelerating the inference step of digitally trained DNNs \cite{Wan2022,Yao2020, Ambrogio2023,Hermes64cores}. However, the increasing computing demands of modern AI models make the training phase orders of magnitude more costly in time and expenses than inference, highlighting the need for efficient hardware acceleration based on the AIMC paradigm. For instance, Gemini 1.0 Ultra required over $5 \cdot 10^{25}$ floating-point operations (FLOPs), approximately 100 days, \(\mathrm{24 \, MW}\) of power, and an estimated cost of 30 million dollars for training \cite{Gemini}. 
\\
\\
Analog training acceleration imposes even more stringent requirements on resistive devices. In addition to inference (i.e., the forward pass), the back-propagation of errors, gradient computation, and weight update steps must be performed during the learning phase. However, in the digital domain updating the weights of a matrix of size NxN requires $O(N^2)$ digital operations, leading to a significant drop in efficiency and speed. Beyond the forward pass, the AIMC approach enables acceleration of (1) backward pass through MVMs transposing the inputs and outputs, (2) gradient computation, and (3) the weight update through gradual bidirectional conductance changes upon external stimuli, all with $O(1)$ time complexity. To achieve this, the ideal analog resistive device should exhibit bidirectional, linear, and symmetric conductance updates in response to an open-loop programming pulse scheme (i.e., without the need for verification following each pulse) \cite{Haensch2019,Woo2018}. 
Promising technologies include redox-based resistive switching memory (ReRAM) \cite{Yin2020,Zahoor2020}, electro-chemical random access memory (ECRAM) \cite{Tang2019}, and capacitive weight elements \cite{Li2018}. Addressing the various non-idealities of these technologies \cite{Ielmini2016} requires the co-optimization of technology and designated training algorithms.
\\
\\
Gokmen et al. \cite{Gokmen2016} proposed an efficient, fully parallel approach that leverages the coincidence of stochastic voltage pulse trains to carry out outer-product calculations and weight updates entirely within memory, in $O(1)$ time complexity. To relax the device symmetry requirements, a novel training algorithm, known as Tiki-Taka, was designed based on this parallel scheme \cite{Gokmen2020}. The primary advantage of the Tiki-Taka approach lies in reduced device symmetry constraints across the entire conductance (G) range, focusing instead on a localized symmetry point where increases and decreases in G are balanced \cite{Gokmen2020}. More recently, the Tiki-Taka version 2 (TTv2) algorithm was demonstrated in hardware \cite{Gong2022} on small-scale tasks using optimized analog ReRAM technology in a 6-Transistor-1ReRAM unit cell crossbar array configuration. However, TTv2 faces some convergence issues when the reference conductance is not programmed with high precision \cite{Rasch2024Agad}. Analog gradient accumulation with dynamic reference (AGAD) learning algorithm (i.e., TTv4) was proposed to overcome the reference conductance limitation, providing enhanced and robust performance \cite{Rasch2024Agad}. 
\\
\\
From a technology perspective, the addition of an engineered conductive-metal-oxide (CMO) layer in a conventional HfO\textsubscript{\textnormal{x}}-based ReRAM metal/insulator/metal (M/I/M) stack has been shown to improve switching characteristics in terms of the number of analog states, stochasticity, symmetry point, and endurance, compared to conventional M/I/M technology \cite{Stecconi2024, Davide_DRC,NHFalcone2024}. However, while CMO/HfO\textsubscript{\textnormal{x}} ReRAM technology has proven to meet all the fundamental device criteria for on-chip training \cite{Stecconi2024}, array-level assessment and BEOL integration remain unexplored. Furthermore, although accelerating DNN training using AIMC is more challenging than inference, a unified technology platform capable of performing on-chip training, retaining the weights, and enabling long-term inference acceleration has yet to be reported. 
\\
\\
This work fills this gap by demonstrating an all-in-one AI accelerator based on CMO/HfO\textsubscript{\textnormal{x}} ReRAM technology, able to perform analog acceleration of both training and long-term inference operations. Such an integrated approach paves the way for highly autonomous, energy-efficient, and continuously adaptable AI systems, opening new paths for real-time learning and inference applications. The flowchart in Fig. \ref{fig1}a illustrates the all-in-one analog training and inference challenge addressed in this study. To achieve this goal, CMO/HfO\textsubscript{\textnormal{x}} ReRAM devices, integrated into the BEOL of a \(\mathrm{130 \, nm}\) complementary metal-oxide-semiconductor (CMOS) technology node with copper interconnects (see "Methods" section "Device fabrication" for details), are arranged in an array architecture using a 1T1R unit cell. Compared to implementations that use multiple transistors to control the resistive switching, the 1T1R unit cell maximizes memory density, which is crucial for storing large AI models on a single chip. Fig. \ref{fig1}b shows an image of the all-in-one analog ReRAM-based AI core used in this work, with the corresponding 8x4 array architecture and the schematic of the BEOL integrated 1T1R cells. The CMO/HfO\textsubscript{\textnormal{x}} ReRAM array is first studied in a quasi-static regime by statistically characterizing the devices' electro-forming step and quasi-static switching response. A physical 3D finite-element model (FEM) is developed to represent the geometry of the conductive filament and analytically describe the charge transport mechanism within these cells. Subsequently, the weight transfer accuracy and conductance relaxation are experimentally characterized on the 8x4 array. These measurements enable the demonstration of the core's inference capabilities, validated through representative MVM accuracy simulations on a 64×64 array. After demonstrating the MVM accuracy of the CMO/HfO\textsubscript{\textnormal{x}} ReRAM core, analog switching experiments using an open-loop identical pulse scheme demonstrated the suitability
of the same core for analog on-chip training acceleration. To assess the training performance, a realistic device model was used in the simulation, accounting for measured characteristics such as non-linear and asymmetric switching behavior, as well as inter- and intra-device variabilities. The training performance was validated using AGAD on fully connected and long short-term memory (LSTM) neural networks, demonstrating scalability from small to large-scale neural networks.

\section{Results}\label{sec2}
\subsection{Quasi-static array characterization and modelling}\label{subsecMemoryCore}
The quasi-static electrical characterization and analytical transport modelling of the 8x4 CMO/HfO\textsubscript{\textnormal{x}} ReRAM array are presented here.
\subsubsection{Filament forming} Fig. \ref{fig2}a shows the current-voltage characteristic of the ReRAM devices in the array, undergoing a soft-dielectric breakdown process, commonly referred to as forming \cite{Padovani2015}. During this step, a quasi-static voltage sweep up to \(\mathrm{3.6 \, V}\) is applied to the top electrode of each ReRAM device, while grounding the source and driving the gate of the corresponding NMOS selector with a constant \(V_\mathrm{G} = \mathrm{1.2 \, V}\) ensuring current compliance. This process leads to the formation of a highly defect-rich conductive filament in the HfO\textsubscript{\textnormal{x}} layer. Due to the high oxygen vacancy ($\rm V_{\rm O}^{\rm \cdot \cdot}$ in Kröger–Vink notation \cite{Kroger}) formation energy, ranging from \(\mathrm{2.8 \, eV}\) to \(\mathrm{4.6 \, eV}\) in HfO\textsubscript{\textnormal{x}} depending on the stoichiometry \cite{Padovani2012,Padovani2013}, defect generation occurs with statistical relevance only during the forming sweep within the HfO\textsubscript{\textnormal{x}} layer \cite{NHFalcone2024}. The subsequent application of a negative voltage sweep up to \(-1.4 \, \mathrm{V}\), with a constant \(V_\mathrm{G} = \mathrm{3.3 \, V}\), induces a radial redistribution of the defects within the CMO layer, consistent with findings in literature \cite{NHFalcone2024}. This process leads to an increase of the ReRAM conductance and is modelled by considering a constant average radius of the conductive filament, with a local electrical conductivity increase of the CMO layer on top of the filament. Refer to the "Methods" section "ReRAM forming modelling" for details. To determine the experimental ReRAM forming voltage, the voltage drop across the NMOS selector must be subtracted from the voltage applied to the 1T1R cell. Fig. \ref{fig2}b shows the experimental transistor output characteristic, from which the resistance in the triode region at \(V_\mathrm{G} = \mathrm{1.2 \, V}\) is measured and used to extract the distribution of \(V_{\mathrm{forming}}^{\mathrm{ReRAM}} \) within the CMO/HfO\textsubscript{\textnormal{x}} ReRAM array (reported in Fig. \ref{fig2}c). Refer to the ”Methods” section ”ReRAM forming voltage extraction” for details. The highly reproducible CMO/HfO\textsubscript{\textnormal{x}} ReRAM forming step exhibits a 100\% yield with a narrow distribution (\(\sigma = \mathrm{75 \, mV}\)) around \(V_{\mathrm{forming}}^{\mathrm{ReRAM}} \approx \mathrm{3.2 \, V}\), making it suitable for integration with \(\mathrm{130 \, nm}\) NMOS transistors rated for \(\mathrm{3.3 \, V}\) operation. 

\subsubsection{Resistive switching and polarity optimization}
\begin{figure}[H]
\centering
\includegraphics[width=1.0\textwidth]{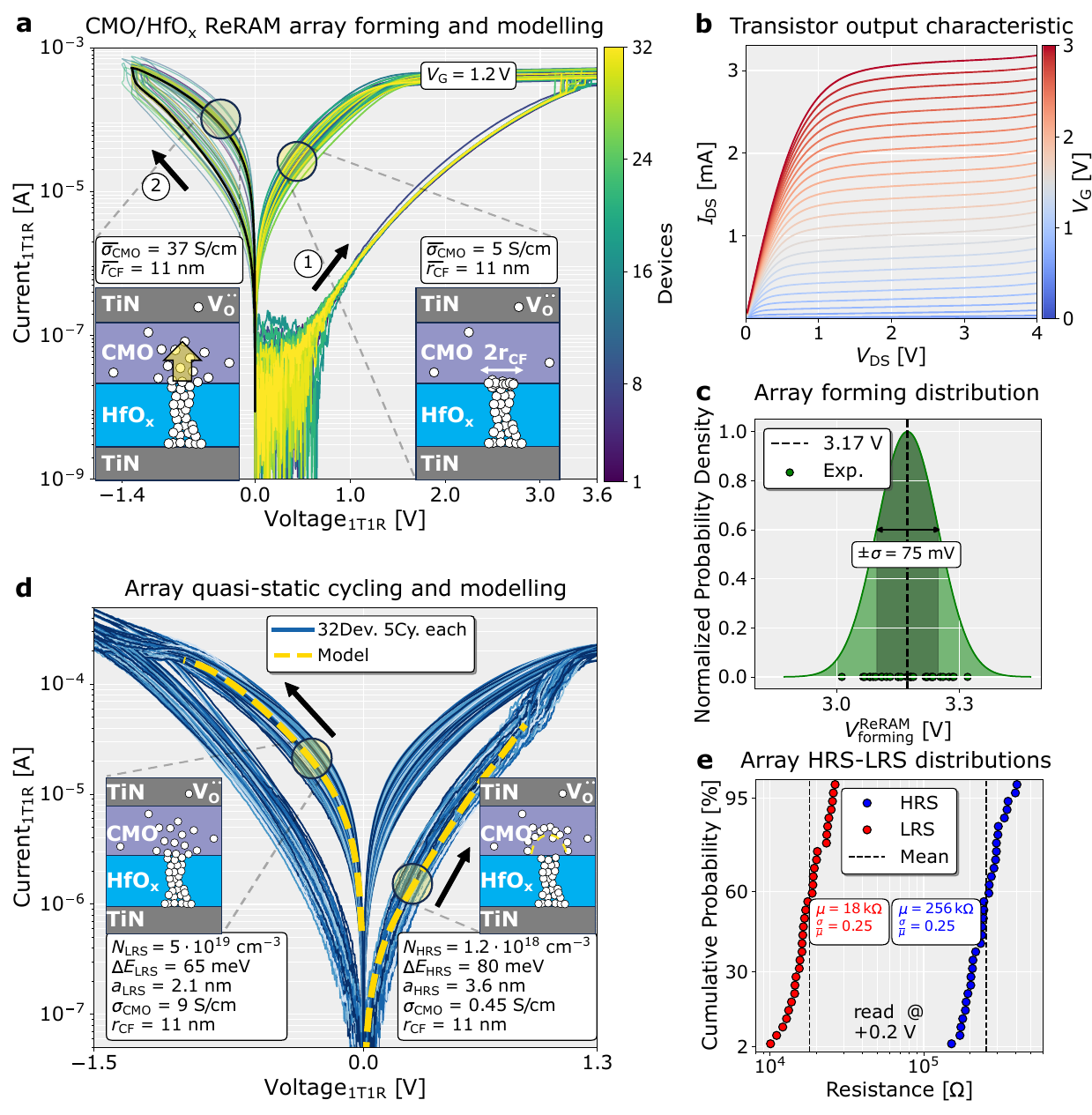}
\caption{ \textbf{ReRAM array quasi-static electrical characterization and modelling.} \textbf{a} (1) Experimental positive forming sweeps (with \(V_\mathrm{G} = \mathrm{1.2 \, V}\)) of the 8x4 CMO/HfO\textsubscript{\textnormal{x}} ReRAM devices in the array. This process results in an average filament radius of \(11\, \mathrm{nm}\) in the HfO\textsubscript{\textnormal{x}} layer. (2) Negative voltage sweeps (with \(V_\mathrm{G} = \mathrm{3.3 \, V}\)) to enable defect redistribution within the CMO layer, resulting in an increase in the conductance of the ReRAM cells. A representative sweep is shown in black. The insets illustrate a schematic representation of the defect arrangement within the stack. \textbf{b} Experimental NMOS transistor output characteristic, with \(V_\mathrm{G}\) up to \(\mathrm{3 \, V}\). \textbf{c} Experimental ReRAM forming voltage distribution measured from the CMO/HfO\textsubscript{\textnormal{x}} ReRAM array. The experimental data used to extract the distribution are represented as green points. \textbf{d} Superposition of 5 I-V quasi-static 8W-cycles (in blue) for each of the 32 devices in the array, using \(V_\mathrm{set} = \mathrm{1.3 \, V}\), \( V_\mathrm{G} = \mathrm{1.1 \, V}\) and \(V_\mathrm{reset} = \mathrm{-1.5 \, V}\), \(V_\mathrm{G} = \mathrm{3.3 \, V}\) for \textit{set} and \textit{reset} processes, respectively. The analytical trap-to-trap tunneling model effectively captures the electron transport in both the LRS and HRS (yellow dashed lines). The physical parameters characterizing the transport, extracted from the model, and a schematic representation of the defect distribution, are presented for both resistive states. \textbf{e} Cumulative probability distributions for both LRS and HRS. For each array cell, the average resistance over 5 I-V cycles in LRS and HRS is defined at a read voltage of \(\mathrm{0.2 \, V}\).}
\label{fig2}
\end{figure}

The underlying physical mechanism behind the resistive switching in analog CMO/HfO\textsubscript{\textnormal{x}} ReRAM devices has been recently unveiled \cite{NHFalcone2024, IMWFalcone2023, ESSDERC24_Galetta}. The current transport is explained by a trap-to-trap tunneling process, and the resistive switching by a modulation of the defect density within the conductive sub-band of the CMO that behaves as electric field and temperature confinement layer. In these works, the analog CMO/HfO\textsubscript{\textnormal{x}} ReRAM device shows a counter-eightwise (C8W) switching polarity, according to the definition proposed in literature \cite{Dittmann2021}. The intrinsically gradual \textit{reset} (from low to high resistance) process, marked by a temperature decrease, occurs during the positive voltage sweep on the ReRAM top electrode, while the exponential \textit{set} (from high to low resistance) process, involving a rapid temperature increase, occurs on the negative side \cite{NHFalcone2024}. However, when arranged in a 1T1R cell configuration based on an NMOS selector, the C8W switching polarity prevents direct control of the transistor's \(V_\mathrm{GS}\) during the exponential \textit{set} process. This results in reduced switching uniformity, which is critical for the array-level adoption of analog CMO/HfO\textsubscript{\textnormal{x}} ReRAM devices. 
\\
\\
For this reason, in this work the analog CMO/HfO\textsubscript{\textnormal{x}} ReRAM devices within the 1T1R cells are optimized to exhibit the desirable 8W switching polarity by extending the current switching model in literature \cite{NHFalcone2024}. To achieve this, following the positive forming and the initial negative voltage sweep, each device in the array is subjected to a forward and backward voltage sweep from 0 to \(-1.5 \, \mathrm{V}\). During this process, oxygen vacancies in the CMO layer radially spread outward, depleting the CMO defect sub-band within a half-spherical volume at the interface with the conductive filament, leading to a \textit{reset} process (Fig. \ref{figS3} in Supplementary Information shows the experimental array's response). Conversely, a voltage sweep from 0 to \(1.3 \, \mathrm{V}\) enables the migration of oxygen vacancies in the CMO layer in the reverse direction, resulting in a \textit{set} transition, controlled by the transistor gate. For each 1T1R cell within the 8x4 array, Fig. \ref{fig2}d shows 5 quasi-static I-V cycling sweeps to experimentally assess the reproducibility of the optimized 8W switching polarity. The electronic transport in both the low-resistive state (LRS) and high-resistive state (HRS) is modelled as a trap-to-trap tunneling process, described by the Mott and Gurney analytical formulation. The physical parameters characterizing the transport in both LRS and HRS ($N_{\rm e}$, $\Delta E_{\rm e}$, $a_{\rm e}$, $\sigma_{\rm CMO}$ and $r_{\rm CF}$) are shown in Fig. \ref{fig2}d. Refer to the ”Methods” section ”Analytical ReRAM transport modelling” for details on the LRS and HRS modelling. Fig. \ref{fig2}e illustrates the cumulative probability distribution of the experimental LRS and HRS within the array, demonstrating device-to-device uniformity and a resistance ratio HRS/LRS of approximately 15, with absolute switching voltages \(\leq \mathrm{1.5 \, V}\). The excellent uniformity of the forming and the optimized 8W-cycling characteristics set the groundwork for AIMC-based inference and training AI-accelerators using the CMO/HfO\textsubscript{\textnormal{x}} ReRAM technology.

\subsection{Analog inference with CMO/HfO\textsubscript{\textnormal{x}} ReRAM core}\label{subsecAnalogInference}
Here, the experimental characterization of the key metrics of the CMO/HfO\textsubscript{\textnormal{x}} ReRAM array relevant to inference performance is presented.
Specifically, the continuous conductance tuning capability is demonstrated over a range spanning approximately one order of magnitude. The trade-off between weight transfer programming noise of CMO/HfO\textsubscript{\textnormal{x}} ReRAM devices and number of required iterations for programming convergence is analyzed across different acceptance ranges. Furthermore, conductance relaxation—defined as the change in conductance over time after programming—is characterized. Finally, the combined impact of weight transfer, conductance relaxation, limited input/output quantization of the digital-to-analog converter (DAC) and analog-to-digital converter (ADC), and IR drop on the array wires is evaluated with respect to MVM accuracy.

\subsubsection{Weight transfer accuracy}\label{subsubsecMultibit}
\begin{figure}[H]
\centering
\includegraphics[width=1.0\textwidth]{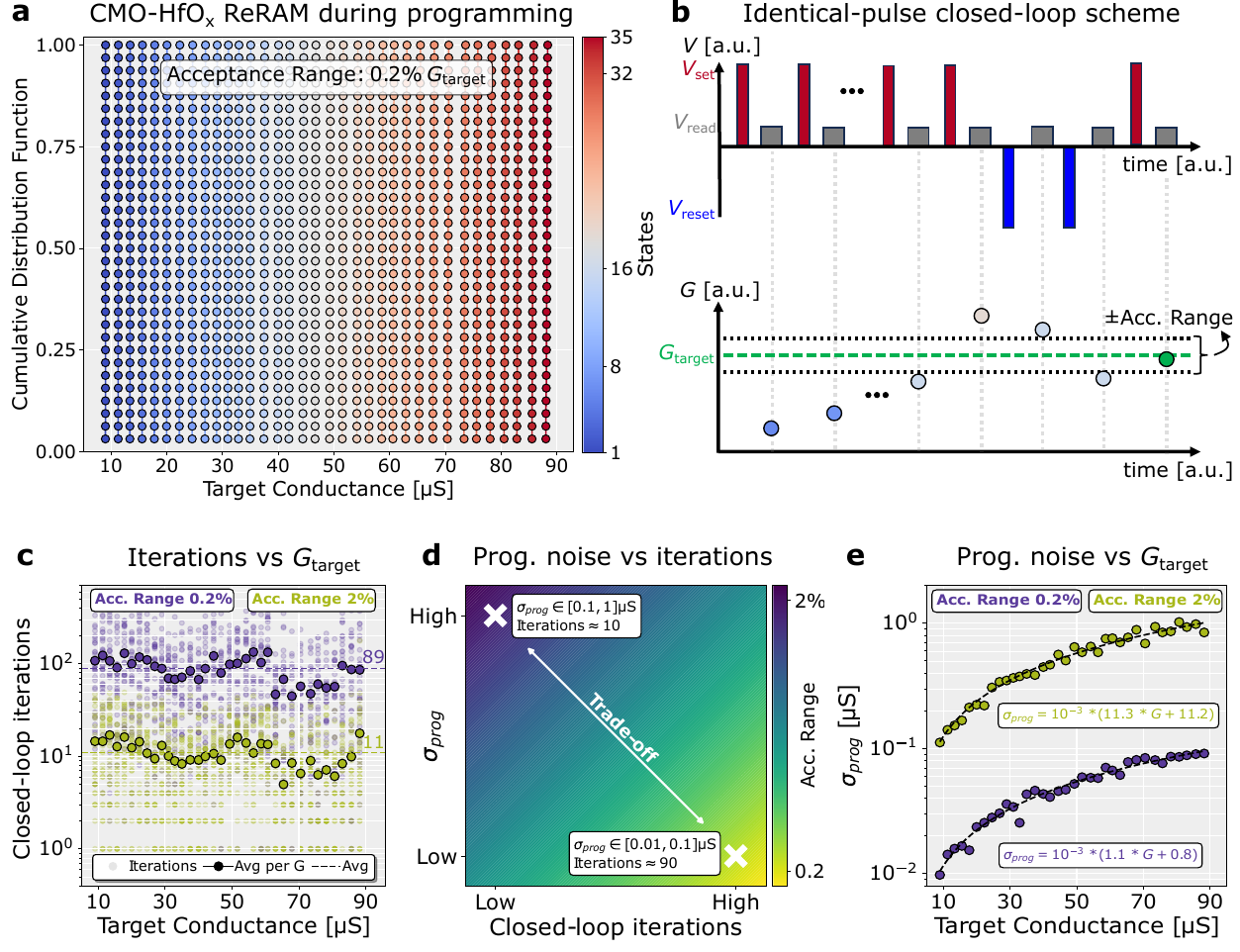}
\caption{\textbf{Weight transfer characterization.} \textbf{a} Cumulative distributions of 35 conductance states obtained using an identical-pulse closed-loop scheme with a 0.2\% \textit{G}\textsubscript{\textnormal{target}} acceptance range. For each distribution, the entire CMO/HfO\textsubscript{\textnormal{x}} ReRAM array was programmed to the corresponding \textit{G}\textsubscript{\textnormal{target}}, and the conductance values measured during the final closed-loop iteration (during programming) is reported. Each dot represents a 1T1R cell. \textbf{b} An example sequence of the identical-pulse closed-loop programming scheme utilized in this work. \textbf{c} Experimental number of closed-loop iterations as a function of \textit{G}\textsubscript{\textnormal{target}} for the two representative acceptance ranges. Each semitransparent point represents a 1T1R cell, the opaque points represent the average number of iterations per \textit{G}\textsubscript{\textnormal{target}}, and the horizontal dashed line indicates the overall average of the opaque points. \textbf{d} Graphical representation of the trade-off between programming noise and the number of iterations required for convergence, as a function of the acceptance range. \textbf{e} Experimental programming noise as a function of \textit{G}\textsubscript{\textnormal{target}} for the two representative acceptance ranges. Each point represents the standard deviation of the normal distribution measured across the entire array. The dashed lines in black indicate the corresponding linear fits.}
\label{fig3}
\end{figure}

In memristor-based AIMC inference accelerators, pre-trained normalized weights are initially mapped into target conductances and subsequently programmed into hardware in an iterative process known as weight transfer. This iterative process, which stops once the programmed conductance converges to the target value within a defined acceptance range, inherently introduces an error due to the analog nature of conductance weights. This error, described by a normal distribution with the standard deviation referred to as programming noise ($\sigma_{\rm prog}$), leads to a drop in MVM accuracy. To quantify this non-ideality, the non-volatile multi-level capability of the CMO/HfO\textsubscript{\textnormal{x}} ReRAM array is characterized. Fig. \ref{fig3}a shows the experimental cumulative distribution of conductance values for 35 representative levels, with all states sharply separated and without any overlap. Fig. \ref{fig3}b shows a schematic representation of the closed-loop (i.e., program-verify) scheme, where identical \textit{set} and \textit{reset} pulse trains are employed to program each ReRAM cell to its target conductance within a desired acceptance range (see ”Methods” section ”Identical-pulse closed-loop scheme” for details). Selecting programming conditions involves a fundamental trade-off: a narrower acceptance range can improve programming precision by reducing programming noise, but it increases the number of iterations required for convergence (see Fig. \ref{fig3}d). Besides the longer programming time, other non-idealities to consider when choosing the acceptance range are (1) the conductance relaxation immediately after programming, which is characterized in \ref{subsubsecRelaxation} for CMO/HfO\textsubscript{\textnormal{x}} ReRAM devices, and (2) read noise, which has already been characterized between 0.2\% and 2\% of \textit{G}\textsubscript{\textnormal{target}} for CMO/HfO\textsubscript{\textnormal{x}} ReRAM devices \cite{Davide_DRC} within a similar conductance range used in this work. 
\\
\\
The trade-off between the programming noise and the number of iterations is characterized for two representative acceptance range intervals: 0.2\% and 2\% of \textit{G}\textsubscript{\textnormal{target}}, respectively. Fig. \ref{fig3}c illustrates the experimental number of pulses needed to converge to the \textit{G}\textsubscript{\textnormal{target}} using the two representative acceptance ranges. On average, each cell requires approximately 11 and 89 \textit{set}/\textit{reset} pulses for acceptance ranges of 2\% and 0.2\% of \textit{G}\textsubscript{\textnormal{target}}, respectively. Since the acceptance range is defined as a percentage of \textit{G}\textsubscript{\textnormal{target}}, the number of iterations required for convergence is almost independent of the target conductance value. In the Supplementary Information, Fig. \ref{figS5}a shows the experimental cumulative distribution of conductance values for the same 35 representative levels presented in Fig. \ref{fig3}a, but using 2\% \textit{G}\textsubscript{\textnormal{target}} as acceptance range. The standard deviation of the representative conductance levels is extracted and fitted as a linear function of the target conductance (dashed lines), as shown in Fig. \ref{fig3}e, for both acceptance ranges. For all conductance levels, a standard deviation of less than 0.1~\textmu S (1~\textmu S) is achieved considering 0.2\% \textit{G}\textsubscript{\textnormal{target}} (2\% \textit{G}\textsubscript{\textnormal{target}}) as the acceptance range. This is more than one order of magnitude lower compared to other memristive technologies, such as phase-change memory (PCM) arrays, targeting similar conductance ranges \cite{Joshi2020, Tsai2019, LeGallo18}. These results demonstrate that CMO/HfO\textsubscript{\textnormal{x}} ReRAM cells achieve an almost ideal weight transfer during programming, enabling the distinction of more than 32 states (5 bits).

\subsubsection{Conductance relaxation and matrix-vector multiplication accuracy}\label{subsubsecRelaxation}
\begin{figure}[H]
\centering
\includegraphics[width=1\textwidth]{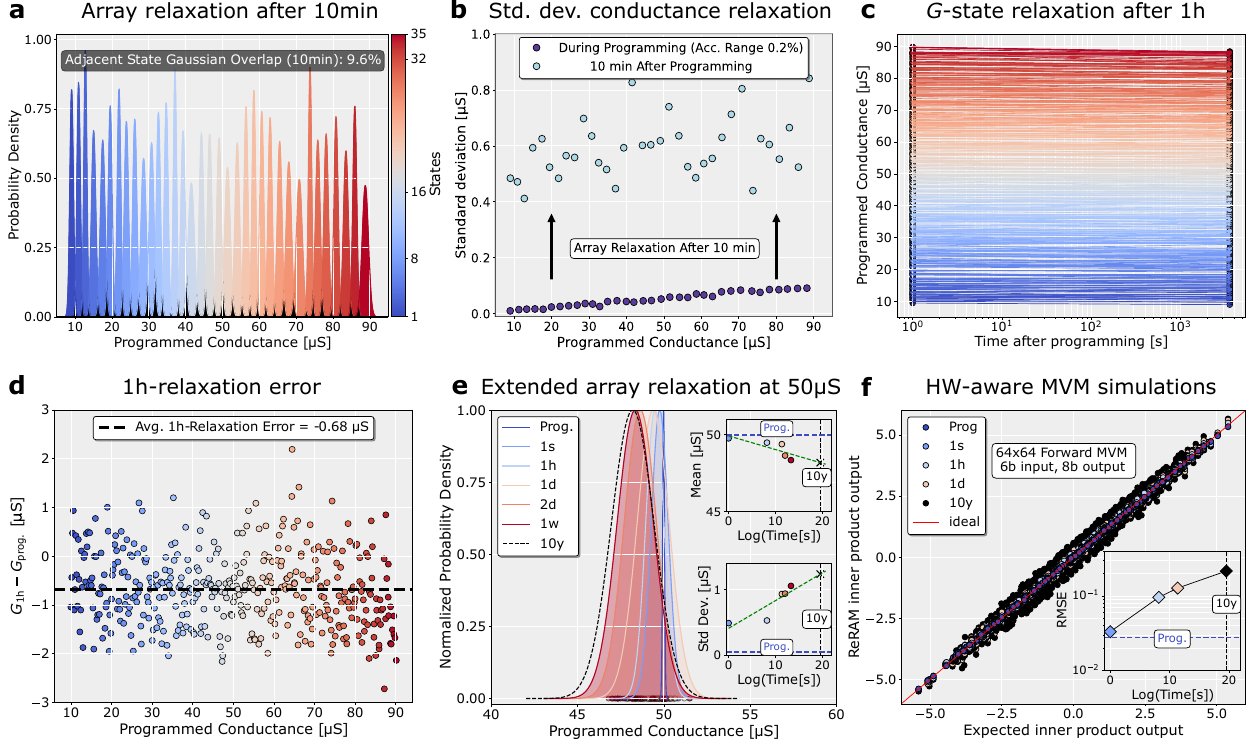}
\caption{\textbf{Conductance relaxation and MVM accuracy.} \textbf{a} Probability density distributions of 35 conductance states approximately 10 minutes after programming. The black areas between adjacent Gaussian distributions represent the overlap of their tails. On average, an overlap of 9.6\% is observed after 10 minutes. \textbf{b} The standard deviations of the 35 conductance states during programming (in purple) and 10 minutes after it (light blue). \textbf{c} Relaxation of 400 conductance states, with one device per G-state, measured 1 second and 1 hour after programming. \textbf{d} Relaxation error 1 hour after programming. A negative and nearly G-independent average error (dashed line) indicates that relaxation in CMO/HfO\textsubscript{\textnormal{x}} ReRAMs tends toward a slight conductance decrease and is state-independent. \textbf{e} Experimental array relaxation of a representative 50 \textmu S state, up to 1 week after programming with 0.2\% \textit{G}\textsubscript{\textnormal{target}} acceptance range. Each probability density distribution is normalized to its maximum for graphical representation. The experimental data used to extract the distributions are represented as points aligned to the y=0 horizontal axis. Insets show the time dependence of the mean and standard deviation. Dashed blue lines represent the conditions during programming, once the convergence to \textit{G}\textsubscript{\textnormal{target}} is reached, while a linear fit (green dashed line) extrapolates the distribution 10 years after programming (dashed black line). \textbf{f} Analog MVM accuracy simulations using a 64x64 CMO/HfO\textsubscript{\textnormal{x}} ReRAM array as a function of time after programming (indicated by different colors). The inset shows the expected RMSE compared to the ideal FP result.  Experimental programming noise, conductance relaxation, limited input/output quantization and IR-drop are considered in this assessment.}
\label{fig4}
\end{figure}

In addition to the excellent weight transfer accuracy during programming as presented in the previous section, the characterization of temporal conductance relaxation is critical to estimate the MVM accuracy over time. In analog ReRAM devices, a significant conductance relaxation has been observed immediately after programming (within 1 second) \cite{Wan2022}. Following this initial abrupt conductance change, the relaxation process slows considerably \cite{Zhao2018,Wan2022}. The physical cause of retention degradation is attributed to the Brownian motion of defects in the resistive switching layer \cite{Zhao2018}. In this section, the conductance relaxation of the CMO/HfO\textsubscript{\textnormal{x}} ReRAM array after programming is characterized. Fig. \ref{fig4}a shows the relaxation of the distributions previously reported in Fig. \ref{fig3}a, approximately 10 minutes after programming. The 35 levels remain distinguishable 10 minutes after programming, with an average overlap of 9.6\% between adjacent states gaussians, while the average standard deviation of the distributions increases to 0.6~\textmu S, showing almost independence from the \textit{G}\textsubscript{\textnormal{target}} (see Fig. \ref{fig4}b). 
\\
\\
The stability of the CMO/HfO\textsubscript{\textnormal{x}} ReRAM conductance states is further assessed on a longer time-scale, up to 1 hour. To achieve so, a linearly spaced \textit{G}\textsubscript{\textnormal{target}} vector within the experimental conductance range of 10~\textmu S to 90 \textmu S is defined, with a fine step of 0.2~\textmu S (400 points). Each \textit{G}\textsubscript{\textnormal{target}} value is programmed into a single ReRAM device within the array. Due to the size mismatch between the array (32 devices) and the \textit{G}\textsubscript{\textnormal{target}} vector (size 400), multiple measurement batches are needed. Fig. \ref{fig4}c shows the experimental relaxation of the 400 programmed states within the entire conductance window, 1 second and 1 hour after programming, executed with the closed-loop scheme (see ”Methods” section
”Identical-pulse closed-loop scheme” for details) and with a 0.2\% \textit{G}\textsubscript{\textnormal{target}} acceptance range. The exhibited conductance error induced by the relaxation process after 1 hour, computed as \(G_{\mathrm{1h}} - G_{\mathrm{prog.}}\), is plotted as a function of the programmed conductances in Fig. \ref{fig4}d. After 1 hour, although both positive and negative relaxation errors are recorded, an average decrease in conductance is observed across all programmed states, with a relaxation error averaging around -0.7~\textmu S. This highlights that the relaxation process in CMO/HfO\textsubscript{\textnormal{x}} ReRAM devices leads, on average, to a decrease in the mean and an increase in the standard deviation of the Gaussian distributions regardless of the initial conductance state. Since the absolute magnitudes of the mean decrease and the standard deviation increase are independent of \textit{G}\textsubscript{\textnormal{target}}, an extended characterization of the relaxation process up to 1 week is conducted for a representative conductance state (50~\textmu S). To achieve this, the array's CMO/HfO\textsubscript{\textnormal{x}} ReRAM devices are programmed using the identical-pulse closed-loop scheme to \textit{G}\textsubscript{\textnormal{target}} of 50~\textmu S, with a 0.2\% \textit{G}\textsubscript{\textnormal{target}} acceptance range. Fig. \ref{fig4}e illustrates the experimental array relaxation over 1 week. The insets display the evolution of both the mean and standard deviation as a function of the logarithm of time after programming (in seconds), using a linear fit to predict the conductance distribution over a 10-year period. 
\\
\\
To assess the accuracy of analog MVM, a comprehensive set of non-idealities—both intrinsic to CMO/HfO\textsubscript{\textnormal{x}} ReRAM devices and at the architecture level—is considered, including finite programming resolution with 0.2\% \textit{G}\textsubscript{\textnormal{target}} acceptance range, conductance relaxation, limited ADC and DAC quantization, and IR-drop across array wires. Fig. \ref{fig4}f shows the hardware-aware simulation results of the analog MVM using CMO/HfO\textsubscript{\textnormal{x}} ReRAM cells, projected for up to 10 years from programming, compared to the expected floating-point (FP) result. The results are generated using a single 64×64 normally distributed random weight matrix and 100 normally distributed input vectors within the range [-1, 1] (see ”Methods” section ”HW-aware simulation of analog MVM” for details). Considering the input and output quantization of 6-bit and 8-bit respectively, the inset illustrates the time evolution of the root-mean-square error (RMSE) of the simulated analog MVM compared to the FP expected result. These results show that the CMO/HfO\textsubscript{\textnormal{x}} ReRAM core enables accurate MVM operations, achieving an RMSE ranging from 0.03 at 1 second to 0.2 at 10 years after programming, compared to the ideal FP case. Fig. \ref{figS6} in the Supplementary Information illustrates the impact of IR-drop and input/output quantization on the RMSE of an MVM performed on a 64×64 array. Over short time scales (within 1 hour), the primary accuracy bottleneck is the limited input/output quantization of 6-bit and 8-bit, respectively. Over longer periods, relaxation effects become the dominant source of non-ideality. In a larger 512×512 array, IR-drop emerges as the main accuracy bottleneck for analog MVM. Compared to the analog ReRAMs studied by Wan et al. \cite{Wan2022}, who report an experimentally determined RMSE of approximately 0.58 under conditions similar to those of this work, CMO/HfO\textsubscript{\textnormal{x}} ReRAMs demonstrate a potential improvement in MVM accuracy by a factor of 20 and 3, 1 second and 10 years after programming, respectively. The excellent MVM accuracy results demonstrate the suitability of CMO/HfO\textsubscript{\textnormal{x}} ReRAM devices for long-term AI inference applications, and lay the foundation for AI training acceleration, where short-term forward and backward MVMs are key steps.

\subsection{Analog training with CMO/HfO\textsubscript{\textnormal{x}} ReRAM core}\label{subsecAnalogTraining}
To efficiently tackle deep learning workloads, the analog AI accelerator must not only perform forward and backward passes (MVMs), but most importantly, allow for weight updates \cite{aihwkit}. During backpropagation, the synaptic weights are modified according to the gradient of the corresponding layer. Therefore, the device conductance must be gradually modified in both positive and negative directions to represent analog weight changes. Analog CMO/HfO\textsubscript{x} ReRAM arrays not only allow for bidirectional conductance updates, but additionally enable parallel weight updating by following a stochastic open-loop pulse scheme \cite{Gokmen2016,Gokmen2020}. Remarkably, the parallel and open-loop update scheme significantly accelerates training compared to serial and closed-loop methods, providing efficiency gains of several orders of magnitude and advantages in system design complexity \cite{Chen2023}. In this section, the bidirectional open-loop response of the CMO/HfO\textsubscript{x} ReRAM array, required during Tiki-Taka training, is characterized. Specifically, the analog conductance potentiation, depression and symmetry point are measured. Subsequently, the devices' responses are statistically reproduced in the open-source 'aihwkit' simulation platform developed by IBM \cite{aihwkit}. Finally, this hardware-aware device model, which includes device variabilities, is used to simulate the training of representative neural networks using the AGAD learning algorithm. This novel analog training algorithm relaxes the symmetry requirements of previous Tiki-Taka versions by incorporating additional digital computations on-the-fly \cite{Rasch2024Agad}.

\subsubsection{Open-loop ReRAM array characterization}\label{openloop_characterization}
\begin{figure}[H]
\centering
\includegraphics[width=1.0\textwidth]{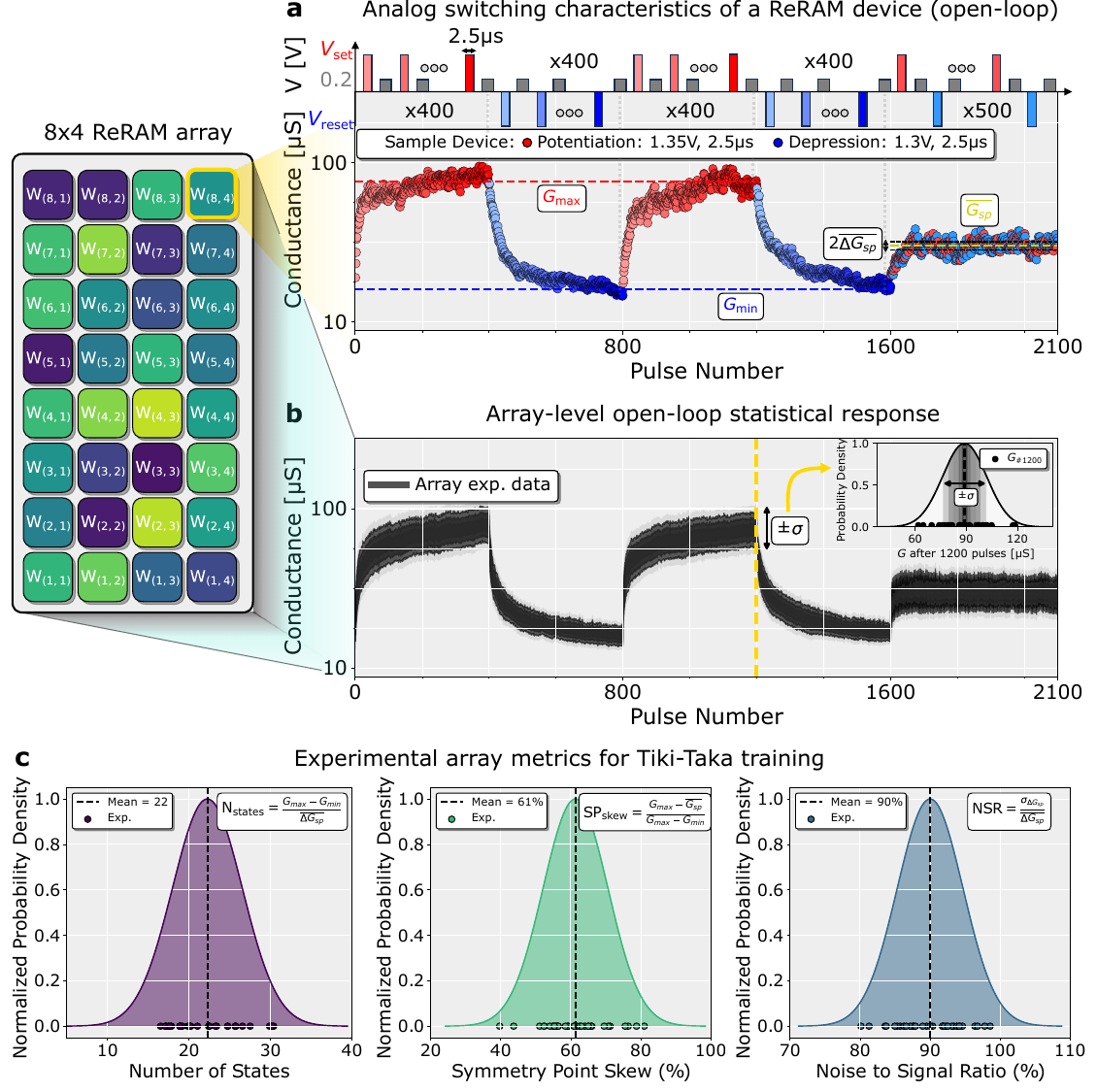}
\caption{\textbf{Open-loop array characterization for on-chip training.} \textbf{a} Bidirectional accumulative response and symmetry point of a representative device in the array. The top inset shows the open-loop identical pulse scheme used for the synaptic potentiation (red) and depression (blue). A conceptual illustration of the 8x4 CMO/HfO\textsubscript{\textnormal{x}} ReRAM array is depicted on the left. \textbf{b} Array statistical open-loop response to identical pulses. The grey area represents the standard deviation of the experimental Gaussian distributions, each corresponding to a specific pulse number. The inset shows a representative example of the experimental G-distribution at pulse number 1200. The raw data can be found in Figure S9 of the Supporting Information. \textbf{c} The experimental probability densities of N$_{\rm states}$, SP$_{\rm skew}$ and NSR, respectively. The experimental data used to extract the distributions are represented as points aligned along the y=0 horizontal axis.}
\label{fig5}
\end{figure}

Fig. \ref{fig5}a shows the experimental conductance change of a representative CMO/HfO\textsubscript{x} ReRAM device within the array upon applying identical-voltage pulse trains with alternating polarity in batches of 400. Subsequently, a sequence of 500 pulses with alternating polarity, consisting of 1-pulse-up followed by 1-pulse-down, is applied to experimentally determine the symmetry point.
The same open-loop programming scheme, with \(V_{\rm set} = 1.35 \, \mathrm{V}\) (\(V_{\rm G} = 1.4 \, \mathrm{V}\)) and \(V_{\rm reset} = -1.3 \, \mathrm{V}\) (\(V_{\rm G} = 3.3 \, \mathrm{V}\)), each lasting 2.5~\textmu s, is applied to all devices in the 8x4 array. The \textit{set}/\textit{reset} pulse width is limited by the experimental setup, although previous work has demonstrated CMO/HfO\textsubscript{\textnormal{x}} ReRAM switching with pulses as short as \(60 \, \mathrm{ns}\) \cite{Davide_DRC}. Due to inter-device (device-to-device) and intra-device (cycle-to-cycle) variabilities, the experimental response of each device to a given number of identical pulses exhibits some level of variability (see Fig. \ref{figS7} in the Supplementary Information). Therefore, for each pulse, a Gaussian distribution of the measured conductance states among the devices is extracted. For statistical relevance, Fig. \ref{fig5}b shows the experimental standard deviation of the array response to the open-loop scheme as a function of the pulse number, represented in grey. To realistically assess the accuracy of analog training with CMO/HfO\textsubscript{x} ReRAM devices, the key figures of merit of the device training characterization—such as the number of states, the symmetry point skew, and the noise-to-signal ratio (NSR)—are first extracted from experimental data, as defined below.%, and then used to feed the device model in the aihwkit.
\begin{align}
   \mathrm{N}_{\rm states} = \frac{G_{\rm max} - G_{\rm min}}{\overline{\Delta G_{\rm sp}}}
   \label{nstates}
\end{align}
\begin{align}
   \mathrm{SP}_{\rm skew} = \frac{G_{\rm max} - \overline{G_{\rm sp}}}{G_{\rm max} - G_{\rm min}}
   \label{skew}
\end{align}
\begin{align}
   \mathrm{NSR} = \frac{ \sigma_{\Delta G_{\rm sp}}  } {\overline{\Delta G_{\rm sp}}}
   \label{NSR}
\end{align}
$G_{\rm max}$ and $G_{\rm min}$ represent the maximum and minimum values extracted from the full conductance swings, while $\overline{G_{\rm sp}}$, $\overline{\Delta G_{\rm sp}}$ and $\sigma_{\Delta G_{sp}}$ denote the values of the mean conductance, mean conductance update and standard deviation of the conductance update at the symmetry point during the 1-pulse-up, 1-pulse-down procedure, respectively. Fig. \ref{fig5}c shows the experimental Gaussian distributions of these metrics for the 32 devices within the array. The results indicate an average of 22 states, with a range from 16 to 33. A shift in the $G_{\rm sp}$ (or SP$_{\rm skew}$) of 61\% is measured, reflecting a negative trend in the device asymmetry where the down response is steeper than the up response. An average NSR of 90\% among the devices is obtained, demonstrating the capability to discriminate between pulses up and down around the symmetry point. This parameter reflects the intrinsic noise on the device's response under identical conditions, highlighting an intra-device variation \cite{aihwkit}. Previous studies on similar CMO/HfO\textsubscript{\textnormal{x}} ReRAM systems \cite{Stecconi2024} extracted these metrics from isolated 1R devices using an optimized open-loop scheme tailored to each device. In contrast, this work demonstrates for the first time that a single open-loop identical pulse scheme enables reliable operation of the entire CMO/HfO\textsubscript{\textnormal{x}} 1T1R array, ensuring consistent performance across the array.

\subsubsection{Tiki-Taka training simulations}\label{subsubsecTT}
\begin{figure}[H]
\centering
\includegraphics[width=1.0\textwidth]{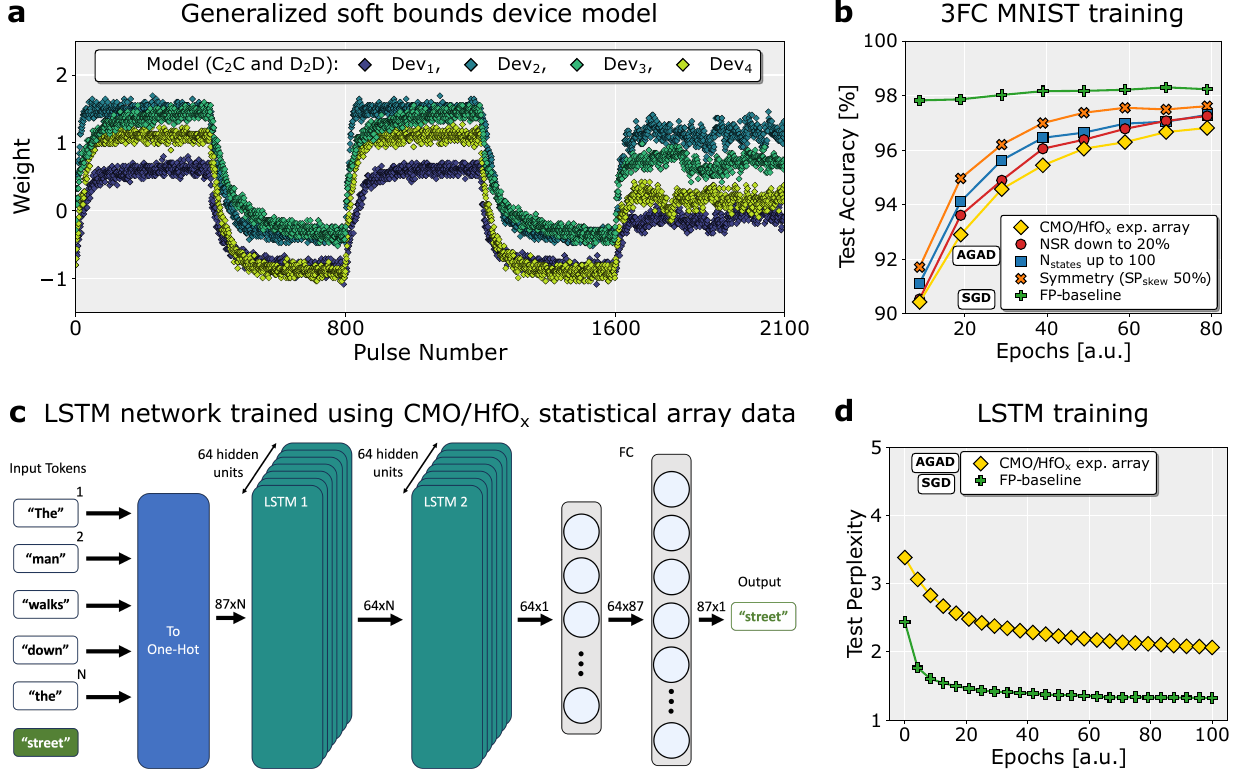} 
\caption{\textbf{Device model and on-chip training simulations.} \textbf{a} Device presets generated using the generalized soft bounds model with experimentally extracted parameters of CMO/HfO\textsubscript{\textnormal{x}}  devices, including inter- and intra-device variabilities. \textbf{b} Training simulations of a 3-layer fully-connected neural network on MNIST (235K parameters), using 32-bit FP precision trained on SGD (in green). Analog training simulations were performed using AGAD considering the empirical distribution of the parameters (in yellow), enhanced NSR (in red), increased N$_{\rm states}$ (in blue), and symmetrical device configurations (in orange). \textbf{c} LSTM network architecture for text forecasting on the \textit{War and Peace} dataset (79K parameters). The architecture  considers a sequence length of 100 tokens and accounts for 2 layers with 64 hidden units. \textbf{d} Training results of the FP baseline (in green) and the analog training with AGAD on the experimental device configuration (in yellow). The training setup can be found in the Supporting Information.}
\label{fig6}
\end{figure}

To perform realistic hardware-aware training simulations, the experimental device response is reproduced on software using the generalized soft bounds model implemented in the 'aihwkit' \cite{Frascaroli2018}, which better captures the bidirectional resistive switching behavior (see Fig. \ref{figS8} in Supplementary Information) and accounts for intra- and inter-device variabilities (see cycle-to-cycle and device-to-device variations in Fig. \ref{fig6}a). Additionally, Gaussian distributions are modelled based on parameters extracted from device characterization ($G_{\rm max}$, $G_{\rm min}$, $\Delta G_{\rm sp}$, NSR, SP$_{\rm skew}$) to account for device-to-device variability observed in the experimental characterization (see "Methods" section "Intra and inter-device variability" for details). This Gaussian fitting approach allows defining various device presets—characterized by the same model but with different parameter settings—to represent the synapses across the neural network. A realistic simulation setup is obtained by exclusively considering experimentally obtained parameters to reproduce the device trace (see "Methods" section "Generalized soft bounds model" for details). The device model is defined based on the observed conductance window and number of states, without assuming asymptotic behavior for an infinite number of pulses. This prevents overestimation of both the conductance window and the number of states (material states), enhancing the fidelity of the simulation.
\\
\\
To validate analog training with CMO/HfO\textsubscript{\textnormal{x}} ReRAM technology, a 3-layer fully connected (FC) neural network was trained on the MNIST dataset for image classification. In addition, the impact of the device's number of states, asymmetry, and noise-to-signal ratio on accuracy and convergence time is evaluated by simulating identical networks in which each property is individually enhanced, while keeping the others fixed at the experimentally derived values. Literature has shown that these device characteristics critically influence the convergence of analog training algorithms \cite{Rasch2024Agad}. Therefore, this method assesses the deviation of the current CMO/HfO\textsubscript{\textnormal{x}} ReRAM device properties from the ideal analog resistive device scenario. Moreover, to show the scalability of the CMO/HfO\textsubscript{\textnormal{x}} ReRAM technology to more computationally-intensive tasks, such as time series processing, a 2-layer long short-term memory (LSTM) network was trained on \textit{War and Peace} text sequences to predict the next token.  
Each network is initially trained using conventional stochastic gradient descent (SGD) based backpropagation with 32-bit FP precision, serving as the baseline performance. Fig. \ref{fig6}b illustrates the accuracy per epoch for the FP-baseline trained with SGD (in green) and the analog network trained using AGAD, evaluated under four different parameter settings: (1) properties extracted from the experimental array (in yellow), (2) reduced NSR to 20\% (in red), (3) average of N$_{\rm states}$ = 100 states (in blue), and (4) zero average device asymmetry (in orange). Using symmetrical device presets, i.e. with an average SP$_{\rm skew}$ of 50\%, improves accuracy by 0.7\% with respect to analog training with CMO/HfO\textsubscript{\textnormal{x}} ReRAM experimentally derived configuration (96.9\%), landing an accuracy of 97.6\%, a 0.7\% lower than the FP-SGD baseline (98.3\%). The other two configurations show less performance improvement, indicating more resilience of the AGAD-training to device's N$_{\rm states}$ and NSR.
\\
\\
Additionally, a 2-layer LSTM network with 64 memory states each (see Fig. \ref{fig6}c), is trained with the experimentally obtained configuration. The performance is measured using the exponential of the cross-entropy loss, i.e. the test perplexity metric, which quantifies the certainty of the token prediction. Results in Fig. \ref{fig6}d demonstrate the capabilities of the CMO/HfO\textsubscript{\textnormal{x}} ReRAM technology on more complex network architectures, such as LSTMs, and computationally demanding tasks, exhibiting performance comparable to the FP-equivalent, with an approximate 0.7\% difference in test perplexity.

\section{Discussion}\label{sec13}
An all-in-one technology platform based on analog filamentary CMO/HfO\textsubscript{\textnormal{x}} ReRAM devices is presented. This platform addresses critical challenges in modern digital AI accelerators by overcoming the physical separation between memory and compute units. It enables the execution of forward and backward MVMs, along with weight updates and gradient computations, directly on a unified analog in-memory platform with $O(1)$ time complexity. This all-in-one approach fundamentally differs from DNN inference-only \cite{Wan2022} and training-only \cite{Stecconi2024,Nate2023} analog accelerators. In inference-only accelerators, DNN weights are trained in software (i.e., off-chip) using traditional digital CPUs or GPUs and then programmed once onto the analog AI hardware accelerator. In training-only accelerators, the long-term retention capabilities and overall MVM accuracy for large array tiles are not assessed. In this work, a novel all-in-one analog computing platform, capable of both on-chip training and inference acceleration, is unveiled.
\\
\\
The CMO/HfO\textsubscript{\textnormal{x}} ReRAM devices are integrated in the BEOL of a NMOS transistor platform in a scalable 1T1R array architecture. The highly reproducible forming step demonstrates compatibility with NMOS rated for \(\mathrm{3.3 \, V}\) operation, while the uniform quasi-static 8W-cycling characteristics, achieved with voltage amplitudes of less than $\pm$ \(\mathrm{1.5 \, V}\), exhibit a significant conductance window and a low off-state. The multi-bit capability of more than 32 states (5 bits), distinguishable after 10 minutes with less than 10\% overlap error, is experimentally demonstrated using an identical-pulse closed-loop scheme. The characterization of the weight transfer reveals record-low programming noise ranging from \(\mathrm{10 \, nS}\) to \(\mathrm{100 \, nS}\), more than one order of magnitude lower than that of other memristive technologies targeting similar conductance ranges \cite{Joshi2020, Tsai2019, LeGallo18}. 
Each conductance distribution exhibits a state-independent relaxation process over time, characterized by a slight shift of the mean toward lower conductance and an increase in the standard deviation. This independence of the relaxation process from the target conductance is advantageous for implementing effective compensation schemes in the future.
\\
\\
Realistic MVM simulations on a 64x64 array tile, considering CMO/HfO\textsubscript{\textnormal{x}} ReRAM device non-idealities such as finite weight transfer resolution, conductance relaxation, limited input/output quantization, and IR-drop across array wires, show an RMSE as low as 0.2 compared to the ideal FP-case, even 10 years after programming. This demonstrates that the CMO/HfO\textsubscript{\textnormal{x}} ReRAM devices improve analog MVM accuracy by a factor of 20 and 3 compared to the state of the art \cite{Wan2022}, 1 second and 10 years after programming, respectively. Although this study was performed at room temperature, previous characterization of a similar CMO/HfO\textsubscript{\textnormal{x}} ReRAM stack demonstrated the thermal stability of the analog states at high temperature (less than 4\% drift after 72 hours at 85~\textdegree C)\cite{Stecconi2024}. Future studies will focus on incorporating the experimental read noise of CMO/HfO\textsubscript{\textnormal{x}} ReRAM devices, characterized between 0.2\% and 2\% of \textit{G}\textsubscript{\textnormal{target}} within a similar conductance range as used in this work \cite{Davide_DRC}, into MVM accuracy simulations. Although read noise is not included in the MVM simulations of this study, no significant additional drop in MVM accuracy is anticipated. In fact, the magnitude of read noise is much smaller than that of the relaxation process and of the effect of reduced input/output quantization, which dominate the RMSE on different timescales. Furthermore, simulation results demonstrate the suitability of CMO/HfO\textsubscript{\textnormal{x}} ReRAM technology for large 512x512 array, with the IR-drop expected to become the primary accuracy bottleneck in this case.
\\
\\
Finally, the electrical response of the CMO/HfO\textsubscript{\textnormal{x}} ReRAM array to an open-loop scheme with identical pulses demonstrates the viability of this technology for on-chip training applications. A realistic device model, accounting for both inter- and intra-device variability, is derived from experimental data. Table \ref{table1} benchmarks the representative device model used in this work on the MNIST dataset against other approaches, highlighting its high fidelity in reproducing experimental device responses. 
\begin{table}[h]
\caption{Device model benchmarking: from simplified approaches to realistic non-ideality modeling}\label{tab1}%
\setlength{\tabcolsep}{2pt}  
\begin{tabular}{@{}p{2.3cm}p{1.8cm}p{2.2cm}p{2.3cm}p{1.4cm}p{1.2cm}p{1.25cm}@{}}
\toprule
ReRAM & Device \space Asymmetry & Analog States & Experimental Data Origin & Algorithm & Model Fidelity & MNIST Accuracy \\
\midrule
Ti/HfO\textsubscript{\textnormal{x}} \cite{Nate2023} & Not-included  & exp. states\footnotemark[2] & BEOL array & TTv2  & Medium & 90.5 \% \\
Ta/TaO\textsubscript{\textnormal{x}} \cite{Nate2023} & Not-included  & exp. states\footnotemark[2] & BEOL array & TTv2 & Medium  & 96.4 \%\\
TaO\textsubscript{\textnormal{x}}/HfO\textsubscript{\textnormal{x}} \cite{Stecconi2024} & included   & material states\footnotemark[3] & Single ReRAMs & TTv2 & Medium & 97.4 \% \\
\textbf{CMO\textsubscript{\textnormal{x}}/HfO\textsubscript{\textnormal{x}}}\footnotemark[1] & \textbf{included}  & \textbf{exp. states}\footnotemark[2] & \textbf{BEOL array} & \textbf{AGAD} & \textbf{High}  & \textbf{96.9 \%} \\
\botrule
\end{tabular}
\footnotetext[1]{\textbf{This work.}}
\footnotetext[2]{Measured number of analog states during open-loop device characterization.}
\footnotetext[3]{The asymptotic number of states under an infinite number of pulses.}
\label{table1}
\end{table}
\\
The impact of the device's number of states, asymmetry and noise-to-signal ratio on training accuracy using the AGAD algorithm on MNIST is evaluated. This analysis demonstrates that, with the current device's experimental properties, AGAD analog training achieves 96.9\% accuracy, comparable to the ideal FP-baseline of 98.3\%. To further improve analog training performance and bring results closer to the software equivalent, the key metric to enhance in the device is the symmetry. Finally, the on-chip analog training capabilities of the CMO/HfO\textsubscript{\textnormal{x}} ReRAM technology are demonstrated on a more complex 2-layer LSTM network, showing comparable performance to its floating-point equivalent.
\\
\\
In conclusion, the novel CMO/HfO\textsubscript{\textnormal{x}} ReRAM all-in-one technology platform presented in this work lays the foundation for efficient and versatile analog chips capable of combining both training and inference capabilities, enabling autonomous, energy-efficient, and adaptable AI systems.

\section{Methods}\label{secMethod}
\subsection{Device fabrication}\label{subsecDeviceFabrication}
The CMO/HfO\textsubscript{\textnormal{x}} ReRAM array is based on 1T1R unit cells. In this configuration, the bottom electrode of the ReRAM device is connected in series to the drain of an n-type metal–oxide–semiconductor (NMOS) selector transistor. The transistor blocks sneak paths and ensures current compliance during electro-forming and programming of the ReRAM device. The NMOS transistors, rated for \(\mathrm{3.3 \, V}\) operation, are fabricated using a standard \(\mathrm{130 \, nm}\) foundry process with copper BEOL interconnects. The ReRAM devices are integrated on metal-8 layer. To prevent the oxidation of the copper vias during the ReRAM stack deposition, the \(\mathrm{70 \, nm}\) thick silicon nitride (SiN\textsubscript{\textnormal{x}}) passivation layer from the foundry is used as a protective layer. On top of that, a \(\mathrm{20 \, nm}\) thick titanium nitride (TiN) bottom electrode and a \(\mathrm{4 \, nm}\) thick hafnium oxide (HfO\textsubscript{\textnormal{x}}) layers are deposited by Plasma-Enhanced Atomic Layer Deposition (PEALD) process at 300~\textdegree C, while maintaining vacuum conditions to avoid oxidation of the TiN layer. Subsequently, a stack of layers consisting of a \(\mathrm{20 \, nm}\) thick conductive metal-oxide (CMO), a \(\mathrm{20 \, nm}\) thick titanium nitride (TiN), and a \(\mathrm{50 \, nm}\) thick tungsten (W) is deposited by sputtering and patterned through a lithography step. A \(\mathrm{100 \, nm}\) thick silicon oxide (SiO\textsubscript{\textnormal{x}}) layer is sputtered as passivation. The passivation layer is then patterned to expose the W top electrode and the copper via in the metal-8 layer beneath the bottom electrode. The ReRAM fabrication is completed using a titanium/gold lift-off process. In this approach, the TiN bottom electrode is connected to the metal-8 via through its vertical sidewalls using gold. The ReRAM BEOL patterning steps are performed through mask-based photolithography performed on a 6$\times$6~mm$^2$ die issued from a Multi Project Wafer (MPW). The area of the CMO/HfO\textsubscript{\textnormal{x}} ReRAM devices presented in this work is 12$\times$12~\textmu m$^2$. Previous studies on CMO/HfO\textsubscript{\textnormal{x}} ReRAM devices have demonstrated scalability down to 200$\times$200~nm$^2$ \cite{Stecconi2024, NHFalcone2024, Davide_DRC}. Due to their filament-type nature, the performance of the ReRAM devices presented in this work is expected to remain similar for smaller areas.

\subsection{ReRAM forming modelling}\label{subsecFormingModelling}
A 3D FEM of the CMO/HfO\textsubscript{\textnormal{x}} ReRAM device, after the forming event, is used to simulate electronic transport by solving the continuity (\ref{Comsol_eq1}) and the Joule-heating (\ref{Comsol_eq2}) equations in steady state: 
\begin{align}
  \nabla \cdot J_{\rm e} = \nabla \cdot (\sigma (-\nabla V) = 0
  \label{Comsol_eq1}
\end{align}
\begin{align}
  \nabla \cdot (- k \nabla T) = J_{\rm e} \cdot E = Q_{\rm e}
   \label{Comsol_eq2}
\end{align}
where $J_{\rm e}$ is the electric current density, $\sigma$ the electrical conductivity, $V$ the electric potential, $k$ the thermal conductivity and $Q_{\rm e}$ the heat source due to Joule heating. From the fit of the experimental array forming data in the low-voltage linear regime (from 0 to \(0.2 \,\mathrm{V}\)), an average filament radius of \(11\, \mathrm{nm}\) is extracted. The electrical and thermal conductivities of the materials in the ReRAM stack are taken from literature \cite{NHFalcone2024}, by considering \( \sigma_{\mathrm{CMO}} = 5 \,\mathrm{S/cm}\) and \( k_{\mathrm{CMO}} = 4 \,\mathrm{W/m K}\) for the CMO layer used in this work. During the subsequent negative voltage sweep, the electrical conductivity of the CMO layer was used as a fitting parameter to model the radial redistribution of defects within the layer. Using experimental array data in the low-voltage linear regime (from 0 to \(\mathrm{-0.2 \, V}\)), the resulting CMO electrical conductivity is \( 37 \,\mathrm{S/cm}\). Fig. \ref{figS1} in Supplementary Information shows the results of the simulations.

\subsection{ReRAM forming voltage extraction}\label{subsecForming}
The forming voltage of each 1T1R cell (\(V_{\mathrm{forming}}^{\mathrm{1T1R}}\)) is defined as the voltage required to trigger the highest current increase (\(\max \left(\frac{dI}{dV}\right)\)) during the quasi-static voltage sweep from 0 to \(3.6 \, \mathrm{V}\) (see Supplementary Information Fig. \ref{figS2}a). The corresponding current is defined as the forming current (\(I_{\mathrm{forming}}^{\mathrm{1T1R}}\)) (see Supplementary Information Fig. \ref{figS2}b). Being the transistor driven by a constant \(V_\mathrm{G} = 1.2 \, \mathrm{V}\), it acts as a series resistor in the triode region before the forming event, when the ReRAM stack is highly insulating. After the forming event, when a conductive filament is created in the ReRAM device, the transistor ensures current compliance in the saturation region. The resistance of the transistor in the triode region at \(V_\mathrm{G} = 1.2 \, \mathrm{V}\) is measured to be \( R_\mathrm{DS} \approx 0.8 \, \mathrm{k\Omega} \) (see Supplementary Information Fig. \ref{figS2}c). Therefore, for each 1T1R cell, the actual ReRAM forming voltage is computed as \(V_{\mathrm{forming}}^{\mathrm{ReRAM}} = V_{\mathrm{forming}}^{\mathrm{1T1R}} - R_{\mathrm{DS}}^{\mathrm{triode}} \cdot I_{\mathrm{forming}}^{\mathrm{1T1R}}\) and reported in Fig. \ref{fig2}c. 

\subsection{Analytical ReRAM transport modelling}\label{subsecTransportModel}
In the 1T1R cell, the electronic current $I_{\rm e}$ is modelled as a trap-to-trap tunneling process within the CMO layer, as described in equation (\ref{I_e_MG}), following the model proposed by Mott and Gurney \cite{Mott_Gurney1950}. This model accounts for electron-hopping conduction across an energy barrier $\Delta E_{\rm e}$, which remains uniform in all directions when there is no electric field applied. However, when an electric field is introduced, it modifies the energy barrier by $\mp$ $ea_{\rm e}E$/2 for forward (backward) jumps, leading to a reduction (increase) in the barrier height. 
\begin{align}
   I_{\rm e}^{\rm Mott-Gurney} = 2 A e a_{\rm e} \nu_{\rm 0,e} N_{\rm e} \exp{(\frac{-\Delta E_{\rm e}}{k_{\rm B} T})} \sinh{(\frac{a_{\rm e}eE}{2k_{\rm B} T})}
   \label{I_e_MG}
\end{align}
In equation \eqref{I_e_MG}, $e$ is the elementary charge, $k_{\rm B}$ is the Boltzmann’s constant, $a_{\rm e}$ is the hopping distance, $\nu_{\rm 0,e}$ is the electron attempt frequency, $N_{\rm e}$ is the density of electronic defect states in the sub-band of the CMO layer, $\Delta E_{\rm e}$ is the zero-field hopping energy barrier, $T$ and $E$ are the local temperature and electric field, respectively, and $A = \rm \pi \it r_{\rm CF}^{\rm 2}$, $r_{\rm CF}$ being the filament radius, is the cross-sectional area of the filament at the interface with the CMO layer. The temperature and electric field in the CMO layer, for both LRS and HRS, are simulated by solving equations (\ref{Comsol_eq1}) and (\ref{Comsol_eq2}), while accounting for the experimental I-V non-linearity (see Supplementary Fig. \ref{figS4} for details). The trap-to-trap tunneling parameters ($N_{\rm e}$, $\Delta E_{\rm e}$, $a_{\rm e}$) are extracted from the fit using the same approach as described in previous works \cite{NHFalcone2024, IMWFalcone2023}.

\subsection{Identical-pulse closed-loop scheme}\label{subsubsecPVInference}
The procedure begins with a quasi-static voltage sweep from 0 to \(-1.5 \, \mathrm{V}\) to \textit{reset} each cell within the array to the HRS. Subsequently, a closed-loop scheme is initiated, which iteratively repeats the following two steps until convergence to \textit{G}\textsubscript{\textnormal{target}} within an acceptance range: (1) read the conductance of the ReRAM cell, and (2) if the measured value is below (above) the target conductance, apply a \textit{set} (\textit{reset}) programming pulse. During this iterative process, the cell conductance may fluctuate multiple times before eventually reaching the acceptance range. Starting from the HRS, this procedure is applied to the CMO/HfO\textsubscript{\textnormal{x}} ReRAM array to sequentially program 35 representative conductance levels, ranging from approximately 10~\textmu S to 90~\textmu S, using acceptance ranges of both 0.2\% \textit{G}\textsubscript{\textnormal{target}} and 2\% \textit{G}\textsubscript{\textnormal{target}}. Unlike the conventional incremental-pulse closed-loop technique previously used for ReRAM \cite{Wan2022, Alibart2012}, where the amplitudes of \textit{set} and \textit{reset} pulses are gradually increased to achieve convergence, this work employs an identical-pulse closed-loop scheme to simplify the pulse generation circuitry design, using only two fixed amplitude values for the \textit{set} (\(1.35 \, \mathrm{V}\) or \(1.5 \, \mathrm{V}\)) and two for the \textit{reset} (\(-1.3 \, \mathrm{V}\) or \(-1.5 \, \mathrm{V}\)) pulses. Specifically, depending on \textit{G}\textsubscript{\textnormal{target}}, three ranges are used: from approximately 10~\textmu S to 30~\textmu S  with \(V_{\rm set} = 1.35 \, \mathrm{V}\) and \(V_{\rm reset} = -1.5 \, \mathrm{V}\); from 30~\textmu S to 60~\textmu S \(V_{\rm set} = 1.35 \, \mathrm{V}\) and \(V_{\rm reset} = -1.3 \, \mathrm{V}\) ; and from 60~\textmu S to 90~\textmu S \(V_{\rm set} = 1.5 \, \mathrm{V}\) and \(V_{\rm reset} = -1.3 \, \mathrm{V}\). Fig. \ref{figS5}b in Supplementary Information shows the flowchart of the identical-pulse closed-loop technique used in this work. The \textit{set}/\textit{reset} pulse width is fixed at 2.5~\textmu s due to setup limitations, even though previous work has demonstrated CMO/HfO\textsubscript{\textnormal{x}} ReRAM switching with pulse width as short as \(60 \, \mathrm{ns}\) \cite{Davide_DRC}. The reading pulse amplitude and width are \(V_{\rm read} = 0.2 \, \mathrm{V}\) and 300~\textmu s, respectively. During the \textit{set}, \textit{reset}, and read operations of each 1T1R cell, the transistor's gate voltage is controlled with constant values of \(V_{\rm G}\) equal to \(1.4 \, \mathrm{V}\), \(3.3 \, \mathrm{V}\), and \(3.3 \, \mathrm{V}\), respectively.

\subsection{HW-aware simulation of analog MVM}\label{subsecInference}
The 'aihwkit' \cite{Rasch2021AFlexible} simulation tool was used to perform MVM assessments including non-ideal behaviors and noise, and their effect on the computation accuracy with respect to floating-point operations. The MVM simulation included the exhibited programming noise, conductance relaxation, input and output quantization, and IR-drop across array wires.
The 'aihwkit' allows to configure such noisy effects for dedicated memristive devices such as PCM by Nandakumar et al. \cite{Nandakumar2019PhaseChange} and ReRAM by Wan et al. \cite{Wan2022}. Therefore, a unique phenomenological noise model for CMO/HfO\textsubscript{\textnormal{x}} ReRAM devices for inference is developed to incorporate into the simulation both the characterized programming noise and conductance relaxation. Additionally, input and output are quantized with 6-bit and 8-bit resolution, respectively, and the IR-drop is considered, with 100~\textmu S as the maximum ReRAM conductance level and a default segment wire resistance of 0.35~$\Omega$. 
\subsubsection{Modelling the programming noise}
For a target conductance  \textit{G}\textsubscript{\textnormal{target}}, the device's programmed conductance is defined as the target value plus normally distributed noise with a standard deviation $\sigma_{\rm prog}$, which is a function of \textit{G}\textsubscript{\textnormal{target}}. As depicted in Fig. \ref{fig3}e, the programming noise ($\sigma_{\rm prog}$) of the CMO/HfO\textsubscript{\textnormal{x}} ReRAM  devices is statistically described by a first-order polynomial equation for a given acceptance range. The polynomial coefficients for acceptance ranges of 2\% and 0.2\% of \textit{G}\textsubscript{\textnormal{target}} are extracted from the characterization and introduced into the simulation environment. To assess the effects of the programming noise, each weight in the normalized matrix (ranging from [-1, 1]) is mapped to its corresponding conductance value (within the range [9, 89]~\textmu S from Fig. \ref{fig3}a), and is then further adjusted by the programming noise described by the extracted linear functions. Therefore, the MVM accuracy can be assessed immediately after programming ($t=0$), see Fig. \ref{fig4}f.

\subsubsection{Modelling the conductance relaxation}
After programming, the conductance levels exhibit relaxation over time, as shown in Fig. \ref{fig4}. Unlike previous ReRAM drift characterizations reported by Wan et al. \cite{Wan2022} the observed relaxation in CMO/HfO\textsubscript{\textnormal{x}} ReRAM is approximately independent of the initial programmed conductance. Consequently, a new modelling approach in the 'aihwkit' is needed to accurately simulate the conductance relaxation effect, which differs from the methods derived from previous literature on ReRAM \cite{Wan2022}. The conductance relaxation mean and standard deviation are modelled independently of \textit{G}\textsubscript{\textnormal{target}} and solely as a function of time after programming. 
The coefficients of the first-order polynomials describing the time dependence of both the mean and standard deviation of the programmed conductance are incorporated into the simulation environment to estimate conductance variations at any given inference time. By doing so, the MVM accuracy can be estimated after a period of time up to 10 years.

\subsection{HW-aware simulation of analog training}\label{devicemodel}
\subsubsection{Generalized soft bounds model}
The generalized soft bounds model (SBM) selection was based on the observed characteristics of the potentiation and depression since the devices did not strictly exhibit thorough saturation at the upper and lower boundaries (see Fig. \ref{figS8} in Supplementary Information). The generalized SBM incorporates a tunable scale exponent ($\gamma$) that describes abrupt and gradual trends toward the maximum and minimum conductance levels. This exponent parameter also varies depending on the conductance update direction. Therefore, the analytical expression of the generalized SBM implemented in the 'aihwkit' includes an asymmetry factor ($\gamma_{\rm up\_down}$) to account for this behavior\cite{aihwkit}. However, these two parameters do not have a direct physical equivalence, and therefore, cannot be derived from experimental traces. Hereby, $\gamma$ and $\gamma_{\rm up\_down}$ are obtained for each device through an independent linear fitting of the generalized SBM to the experimental response. In addition to the analytical parameters of the generalized SBM, devices in the 'aihwkit' are defined by a set of parameters that can be extracted from experimental traces. More precisely, the empirical maximum and minimum conductance, minimum conductance step size and its standard deviation, and the asymmetry between up and down response are considered ($G_{\rm max}$, $G_{\rm min}$, $\Delta G_{\rm sp}$, $\sigma_{\Delta G_{\rm sp}}$, and $up\_down$). More details on the $up\_down$ parameter are provided in the Supplementary Information. In this regard, each simulated device is defined by 6 parameters: four empirically obtained ($G_{\rm max}$, $G_{\rm min}$, $\Delta G_{\rm sp}$ and $up\_down$) and two analytically modelled from SBM linear fitting ($\gamma$ and $\gamma_{\rm up\_down}$).

\subsubsection{Intra and inter-device variability}
By extracting the standard deviation of the minimum conductance step size ($\sigma_{\Delta G_{\rm sp}}$) from the experimental traces and incorporating it into the simulation's device model, the device response intrinsically includes noise from cycle to cycle. This provides a realistic device behavior with intra-device variability. Furthermore, the network devices shall include inter-device variabilities to perform physically accurate simulations. To achieve this, two multi-variate Gaussian distributions, G$_{\rm 1}$ and G$_{\rm 2}$, are created (see Fig. \ref{figS9} in Supplementary Information). G$_{\rm 1}$ is extracted from the experimentally obtained parameters: N$_{\rm states}$ (which accounts for variations across devices in the G-range and step) and SP in the normalized G-range, whereas G$_{\rm 2}$ is fitted to the analytical model parameters obtained from the fitted generalized SBM ($\gamma$ and $\gamma_{\rm up\_down}$). Therefore, variables from G$_{\rm 1}$ showed statistical independence from those of G$_{\rm 2}$. New device instances are independently sampled from the two Gaussian distributions to represent synapses on the DNN layers. The instantiated CMO/HfO\textsubscript{\textnormal{x}} ReRAM devices include variations in the device response, conductance ranges, and asymmetrical behavior, thus providing a more hardware-aware and realistic scenario for analog training simulation. 

% .bib or .bbl
%\bibliography{sn-bibliography}

\backmatter

\bmhead{Supplementary information}
This manuscript is supported by additional supplementary information provided in a separate document.

\bmhead{Acknowledgements}
The authors acknowledge the Binnig and Rohrer Nanotechnology Center (BRNC) at IBM Research Europe - Zurich. Special thanks go to Jean-Michel Portal, Eloi Muhr and Dominique Drouin for their contributions to the design of the NMOS transistors used in this work. The authors also extend their gratitude to Stephan Menzel for the fruitful discussions and to Ralph Heller for his assistance in wire-bonding the chip. This work is funded by SNSF \textit{ALMOND} (grantID: 198612), by the European Union and Swiss state secretariat SERI within the H2020 \textit{MeM-Scales} (grantID: 871371), \textit{MANIC} (grantID: 861153), \textit{PHASTRAC} (grantID: 101092096) and \textit{CHIST-ERA UNICO} (20CH21-186952) projects.

\bmhead{Author contributions}
Conceptualization: D. F. F. and V. B.; hardware fabrication: D. F. F. and L. B. L.; electrical characterization: D. F. F, W. C., T. S., F. H., physical simulations: D. F. F.; inference and training simulations: V. C., D. F. F.; NMOS transistor design : N. G., F. A.; result interpretation: D. F. F., V. C., W. C., V. B., M. G., A. L. P. and B. J. O., supervision: V. B. and B. J. O.; manuscript writing: D. F. F., V. C.; data curation: D. F. F., V. C. and V. B.; manuscript review and editing: all authors; funding acquisition: B. J. O. and V. B.

\bmhead{Competing interests}
The authors declare no competing interests.

\bmhead{Data availability}
The data that support the plots within this paper and other findings of this study are available from the corresponding authors upon reasonable request.

\bmhead{Code availability}
The repositories containing the source codes used in this work for analog inference simulations and CMO/HfO\textsubscript{\textnormal{x}} ReRAM noise model can be found at \href{https://github.com/IBM/aihwkit/blob/master/examples/34_ReRAM_inference.py}{this link} and \href{https://github.com/IBM/aihwkit/blob/master/src/aihwkit/inference/noise/reram.py}{this link}, respectively.

%\appendix
%\renewcommand{\thefigure}{S\arabic{figure}} 
%\setcounter{figure}{0} 

\appendix
\renewcommand{\thefigure}{S\arabic{figure}} 
\setcounter{figure}{0} % Reset della numerazione delle figure

\section*{Supplementary Information}%\label{secA1}
\begin{figure}[H]
\centering
\includegraphics[width=1.0\textwidth]{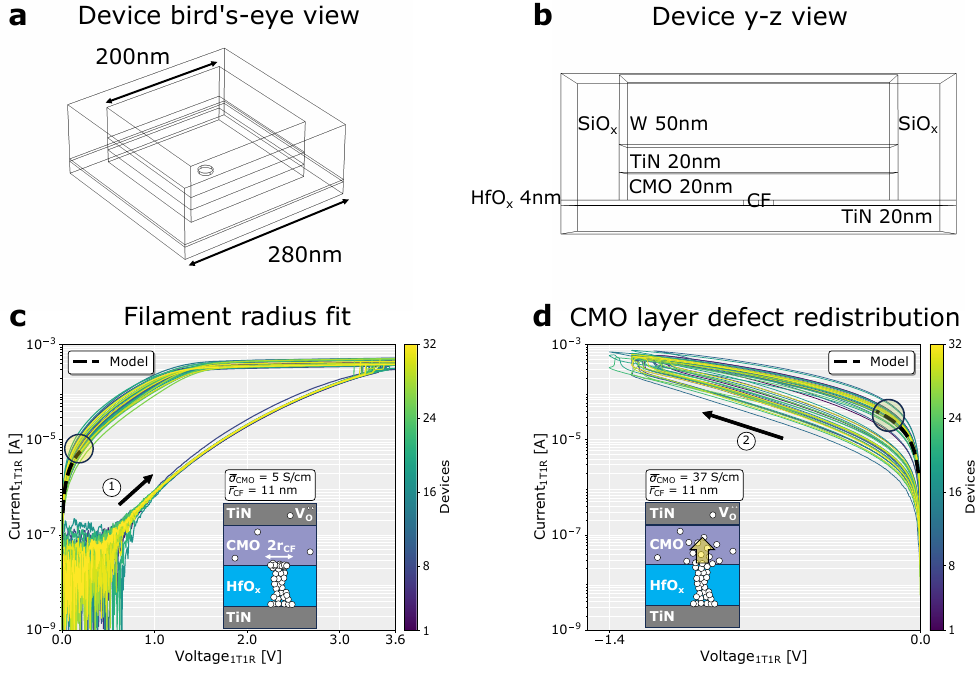}
\caption{\textbf{ReRAM forming modelling.} The CMO/HfO\textsubscript{\textnormal{x}} ReRAM device is simulated using a 3D FEM in COMSOL Multiphysics 5.2 software. \textbf{a} The bird’s-eye view and \textbf{b} the lateral y-z view of the device’s geometry and material stack are shown. Due to the temperature and electric field confinement, an effective device area of 200$\times$200~nm$^2$ is considered for the simulation to reduce computational resource demands. \textbf{c} The experimental array forming data in the low-voltage linear regime (from 0 to \(0.2 \,\mathrm{V}\)) are fitted to extract the average filament radius. \textbf{d} The increase in experimental conductance resulting from a negative voltage sweep after the forming event is modelled as an effective increase in the electrical conductivity of the CMO layer, due to a radial redistribution of defects.}
\label{figS1}
\end{figure}

\begin{figure}[H]
\centering
\includegraphics[width=1.0\textwidth]{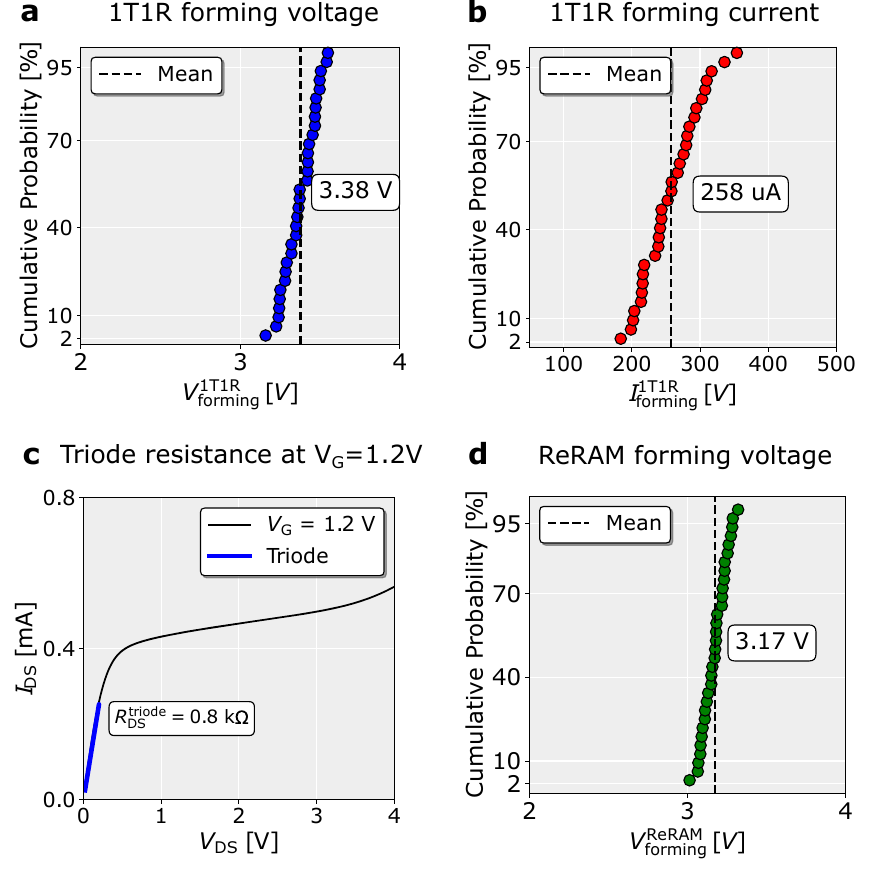}
\caption{\textbf{Experimental CMO/HfO\textsubscript{\textnormal{x}} ReRAM array forming data.} \textbf{a} The forming voltage distribution of the 1T1R cells within the array, defined as the voltage required to trigger the highest current increase during the quasi-static voltage sweep from 0 to \(3.6 \,\mathrm{V}\) in Fig. \ref{fig2}a of the manuscript. \textbf{b} The array forming current distribution corresponding to \( V = V_{\mathrm{forming}}^{\mathrm{1T1R}} \). \textbf{c} The experimental resistance of the transistor in the triode region at \(V_\mathrm{G} = \mathrm{1.2 \, V}\), extracted from a linear fit from 0 to \(0.2 \,\mathrm{V}\) of the transistor output characteristic. \textbf{d} The forming voltage distribution of the ReRAM array, shown in Fig. \ref{fig2}c of the manuscript, computed as \(V_{\mathrm{forming}}^{\mathrm{ReRAM}} \)=  \(V_{\mathrm{forming}}^{\mathrm{1T1R}} \) - \(R_{\mathrm{DS}}^{\mathrm{triode}} \)* \(I_{\mathrm{forming}}^{\mathrm{1T1R}} \).}
\label{figS2}
\end{figure}

\begin{figure}[H]
\centering
\includegraphics[width=0.6\textwidth]{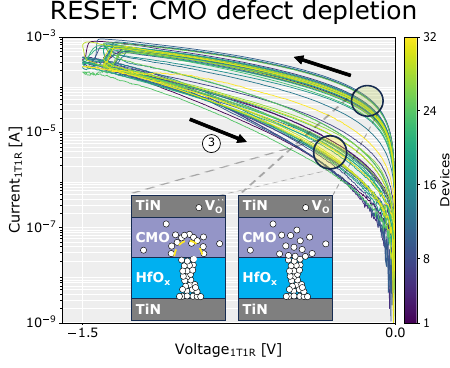}
\caption{The experimental array’s response to the voltage sweep from 0 to \(-1.5 \,\mathrm{V}\), following the positive forming and the initial negative voltage sweep (denoted as step (1) and (2) in Fig. \ref{fig2}a of the manuscript, respectively). The oxygen vacancies in the CMO layer radially spread outward, depleting the CMO defect sub-band within a half-spherical volume at the interface with the conductive filament, leading to a \textit{reset} process.}
\label{figS3}
\end{figure}

\begin{figure}[H]
\centering
\includegraphics[width=1.0\textwidth]{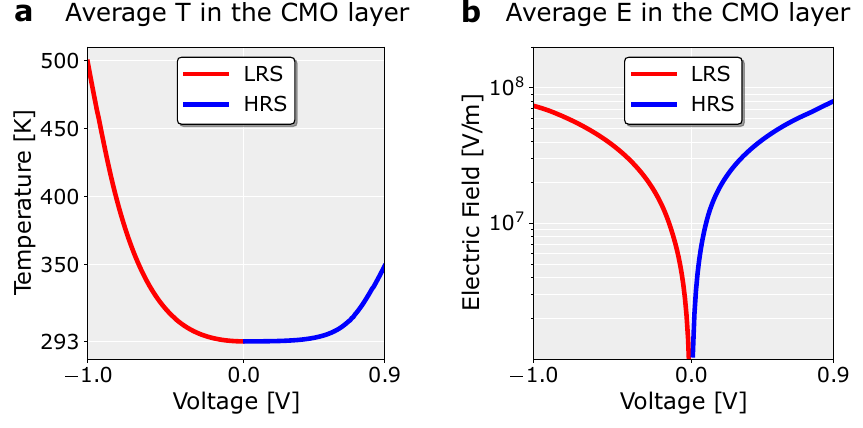}
\caption{
The voltage-dependent evolution of \textbf{a} the average temperature and \textbf{b} electric field within a 3D half-spherical volume of the CMO layer situated atop the conductive filament in both HRS and LRS is presented. These trends serve as inputs for equation (6) of the manuscript.}
\label{figS4}
\end{figure}

\begin{figure}[H]
\centering
\includegraphics[width=1.0\textwidth]{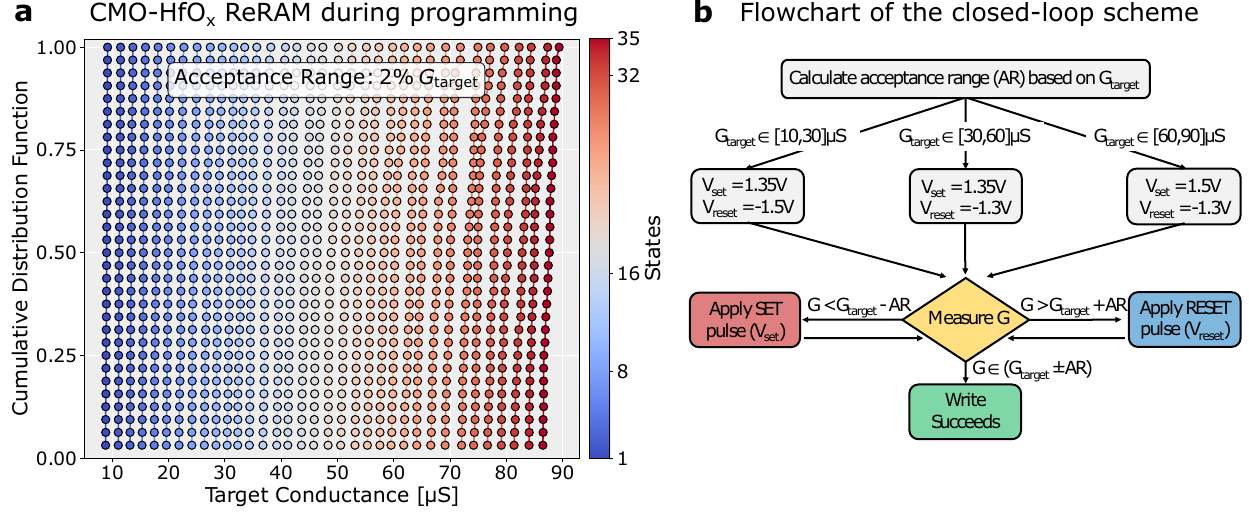}
\caption{\textbf{a} The experimental cumulative distribution of conductance values for 35 representative programmed levels using 2\% of \textit{G}\textsubscript{\textnormal{target}} as acceptance range. The closed-loop scheme based on identical pulses shown in Fig. \ref{fig3}b of the manuscript and detailed in Methods is used. \textbf{b} Flowchart illustrating the identical-pulse closed-loop technique used for programming the ReRAM array into target analog conductance range.}
\label{figS5}
\end{figure}

\begin{figure}[H]
\centering
\includegraphics[width=1.0\textwidth]{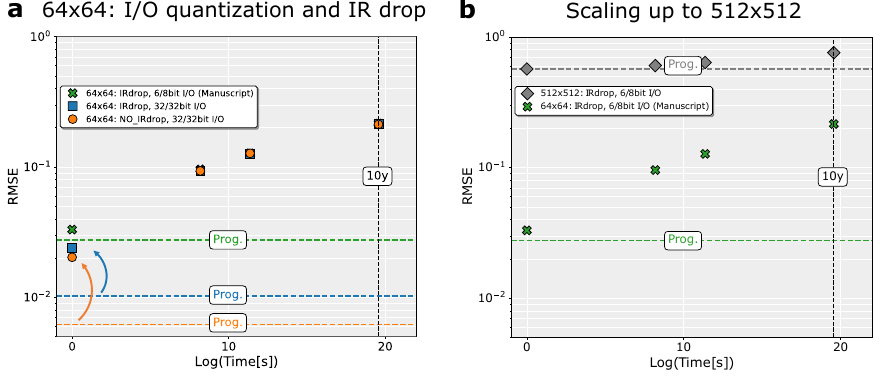}
\caption{ \textbf{The individual impact of IR-drop across array wires and input/output bit quantization on MVM accuracy.} \textbf{a} Simulated RMSE compared to FP ideal results using 64x64 analog CMO/HfOx ReRAM array, shown as a function of the time after programming. Dashed horizontal lines represent the RMSE during programming, considering programming noise (with 0.2\% \textit{G}\textsubscript{\textnormal{target}} as the acceptance range) but excluding relaxation effects. With 32-bit input/output quantization and no IR-drop (orange dots), an RMSE as low as 6 $10^{-3}$ is achieved during programming, which immediately increases (see the arrow) after relaxation (within \(\mathrm{1 \, s}\)). Including the realistic IR-drop results in an overall RMSE increase (blue squares). Finally, reducing input/output quantization to 6/8 bits, respectively, leads to a further accuracy loss (green crosses), demonstrating that at short timescales (within 1 hour), the main analog MVM accuracy bottleneck is the reduced input/output quantization. After 1 hour, all cases converge, showing that the accuracy bottleneck is then dominated by the relaxation process. \textbf{b} By scaling up to a 512x512 array size (grey diamonds) and considering input/output quantization of 6/8 bits, IR-drop emerges as the primary bottleneck for analog MVM accuracy.}
\label{figS6}
\end{figure}

\begin{figure}[H]
\centering
\includegraphics[width=1.0\textwidth]{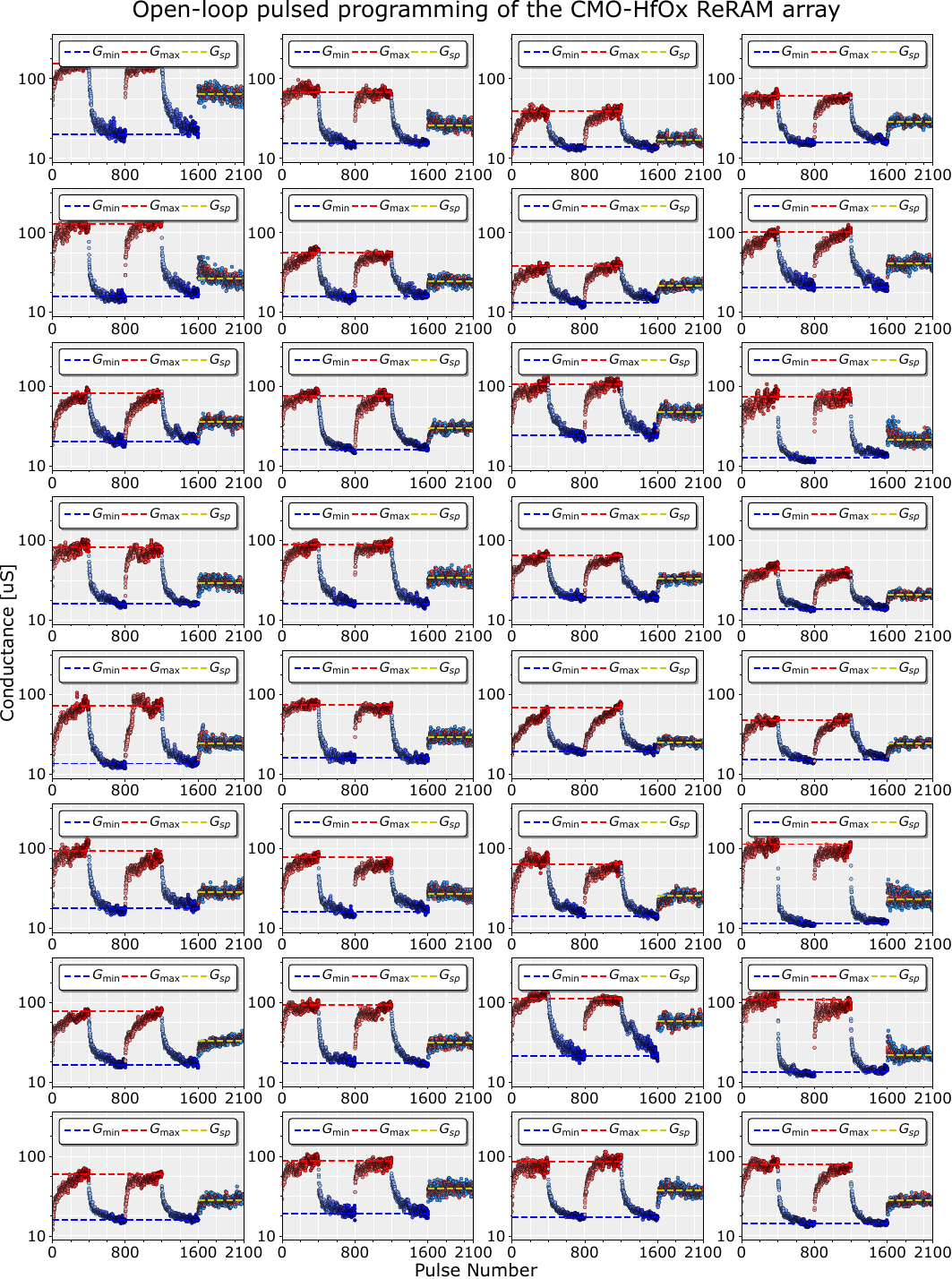}
\caption{The experimental response of the 8x4 CMO/HfO\textsubscript{\textnormal{x}} ReRAM devices within the array to the open-loop programming pulse scheme (shown in Fig. \ref{fig5}b of the manuscript) is shown. The \textit{set} and \textit{reset} pulse amplitudes are \(1.35 \,\mathrm{V}\) (\(V_\mathrm{G} = \mathrm{1.4 \, V}\)) and \(-1.3 \,\mathrm{V}\) (\(V_\mathrm{G} = \mathrm{3.3 \, V}\)), respectively, with a constant width of 2.5~\textmu s due to setup limitations.}
\label{figS7}
\end{figure}

\begin{figure}[H]
\centering
\includegraphics[width=1.0\textwidth]{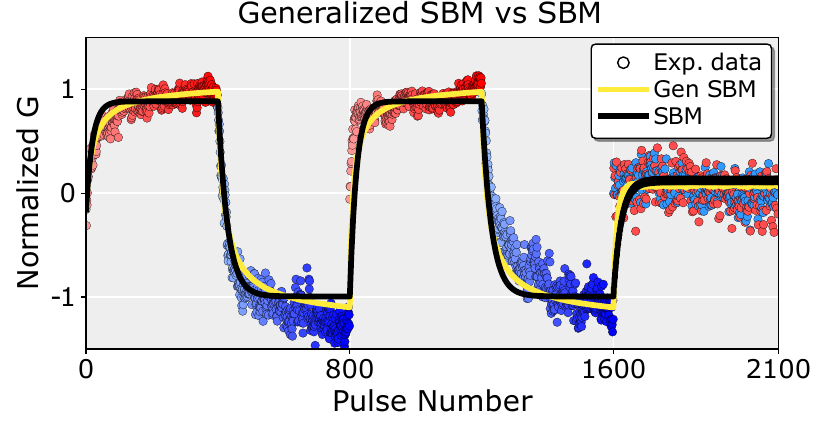}
\caption{The experimental open-loop pulsed response of a representative CMO/HfO\textsubscript{\textnormal{x}} ReRAM device within the array shows that the potentiation and depression characteristics do not inherently saturate at the upper and lower boundaries. The generalized soft bounds model (yellow line) better captures this experimental trend compared to the saturated soft bounds model (black line).}
\label{figS8}
\end{figure}

\begin{figure}[H]
\centering
\includegraphics[width=1.0\textwidth]{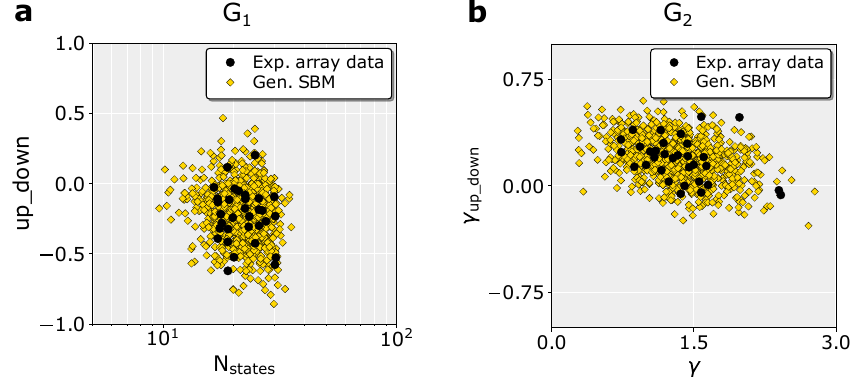}
\caption{
Multi-variate Gaussian distributions to reproduce the experimental inter-device variability. \textbf{a} Multi-variate gaussian G1 distribution of the experimental number of states and device asymmetry ($up\_down$). \textbf{b} Gaussian G2 distribution of the analytical parameters $\gamma$ and $\gamma_{\rm up\_down}$ extracted from the generalized soft bounds model fitting to the experimental traces.}
\label{figS9}
\end{figure}

\subsection*{Device modelling}
\subsubsection*{$up\_down$ parameter}
The $up\_down$ parameter is defined for the generalized soft bounds model in the simulation environment of the ‘aihwkit’ as the directional bias between the up and down update size ($\Delta G^+$ and $\Delta G^-$). In addition, the minimum step in each direction \textit{d} is described by the following expression \cite{Rasch2021AFlexible}.
\begin{align}
    \Delta G^d = \Delta G_{SP} (1 + d\beta + \sigma_{d-to-d})
   \label{eqS1}
\end{align}
where \textit{d} is -1 or 1 depending on the update direction. In contrast, the symmetry point is defined for each device as follows \cite{Rasch2024Agad}:
\begin{align}
    SP = [\Delta G^+ - \Delta G^-]/[\Delta G^+ / (b_{\rm max} - \Delta G^+/b_{\rm min})]
   \label{eqS2}
\end{align}
Where $\Delta G^+$, $\Delta G^-$ define the minimum step size in the up and down direction respectively, and $b_{\rm max}$ and $b_{\rm min}$ represent the upper and lower bounds of the conductance. Therefore, considering an independent definition of each device (i.e. zero d-to-d variability) and a normalized conductance range between -1 and 1, the symmetry point device-level characteristic and the $up\_down$ analytical parameter are equivalent.

\subsubsection*{Training setup}
For result replicability, the experimental parameters are incorporated into the simulation environment, where the Noise-to-Signal Ratio (NSR) is represented by  'dw\_min\_std', normalized SP by ‘up\_down’, normalized maximum and minimum conductances by ‘w\_max’ and ‘w\_min’ and min conductance step by ‘dw\_min’. 
\\
\\
From this device model, analog training simulations were performed using AGAD considering a learning rate to update the weights of 1e-2, ‘fast\_lr’ of 0.1 to update matrix, ‘transfer\_every’ 3 iterations and batch size of 32. The FP baseline was obtained with SGD training using a learning rate of 1e-3 and batch size of 32.

\end{document}